\documentclass[letterpaper]{article} 
\usepackage{aaai2026}  
\usepackage{times}  
\usepackage{helvet}  
\usepackage{courier}  
\usepackage[hyphens]{url}  
\usepackage{graphicx} 
\usepackage{natbib}  
\usepackage{caption} 
\usepackage{algorithm}
\usepackage{algorithmic}

\usepackage{amsmath, amssymb}
\usepackage{subcaption}
\usepackage{booktabs}       
\usepackage[table]{xcolor}
\usepackage{enumitem}
\usepackage{bibunits}
\usepackage{dsfont}
\usepackage{newfloat}
\usepackage{listings}
\DeclareCaptionStyle{ruled}{labelfont=normalfont,labelsep=colon,strut=off} 
\floatstyle{ruled}
\newfloat{listing}{tb}{lst}{}
\floatname{listing}{Listing}
\title{Tokenize Once, Recommend Anywhere: Unified Item Tokenization for Multi-domain LLM-based Recommendation}
\author{
    Yu Hou\textsuperscript{\rm 1},
    Won-Yong Shin\textsuperscript{\rm 1}\thanks{Corresponding Author.}
}
\affiliations{
    \textsuperscript{\rm 1}Yonsei University\\

    \{houyu, wy.shin\}@yonsei.ac.kr
}

\begin{document}

\maketitle

\begin{abstract}
Large language model (LLM)-based recommender systems have achieved high-quality performance by bridging the discrepancy between the item space and the language space through item tokenization. However, existing item tokenization methods typically require training separate models for each item domain, limiting generalization. Moreover, the diverse distributions and semantics across item domains make it difficult to construct a unified tokenization that preserves domain-specific information. To address these challenges, we propose {\bf \textsf{UniTok}}, a {\bf Uni}fied item {\bf Tok}enization framework that integrates our own mixture-of-experts (MoE) architecture with a series of codebooks to convert items into discrete tokens, enabling scalable tokenization while preserving semantic information across {\it multiple item domains}. Specifically, items from different domains are first projected into a unified latent space through a shared encoder. They are then routed to {\it domain-specific} experts to capture the {\it unique} semantics, while a {\it shared} expert, which is always active, encodes common knowledge transferable across domains. Additionally, to mitigate {\it semantic imbalance} across domains, we present a mutual information calibration mechanism, which guides the model towards retaining similar levels of semantic information for each domain. Comprehensive experiments on wide-ranging real-world datasets demonstrate that the proposed \textsf{UniTok} framework is {\bf (a) highly effective:} achieving up to 51.89\% improvements over strong benchmarks, {\bf (b) theoretically sound:} showing the analytical validity of our architectural design and optimization; and {\bf (c) highly generalizable:} demonstrating robust performance across diverse domains without requiring per-domain retraining, a capability not supported by existing baselines.
\end{abstract}

\begin{links}
    \link{Code}{https://github.com/jackfrost168/UniTok}
\end{links}


\section{Introduction}

Large language models (LLMs) have recently become a promising paradigm for generative recommendation \cite{rajput2023recommender, hua2023index}, leveraging their strong generalization, language understanding, and world knowledge to support personalized recommendation beyond traditional language processing tasks. To effectively use LLMs for recommendation, items must be indexed using identifiers, a process known as item tokenization \cite{rajput2023recommender}. Item tokenization converts items into {\it discrete tokens}, such as ID-based representations~\cite{hua2023index}, textual descriptors~\cite{zhang2021language}, or codebook-based identifiers~\cite{rajput2023recommender}. This bridges the gap between the item space and the language space, enabling LLMs to process items as part of natural language sequences and making generative recommendations feasible. 

Existing item tokenization methods \cite{rajput2023recommender, wang2024learnable} are primarily tailored to items in single-domain settings, necessitating the training of separate tokenizers for each item domain (hereafter, “domain” refers to “item domain” for brevity). In practice, this domain-specific design aligns with the fact that recommender systems are often deployed independently per domain; thus, users rarely perceive quality issues. However, as recommendation tasks increasingly span multiple domains, such as diverse item categories or services, this siloed approach leads to inefficiencies in training, deployment, and maintenance, ultimately hindering scalability. In contrast, other machine learning fields have seen a growing shift towards building {\it unified models} for multi-domain learning, driven by the need to reduce redundant training, improve parameter efficiency, and facilitate knowledge sharing across domains. Notable advances in this direction have been achieved in language processing \cite{GururanganMSLBD20, raffel2020exploring} and computer vision \cite{ullah2022meta, jain2023damex}, demonstrating the feasibility and value of such generalization.

\begin{figure}[t]
    \centering
    \begin{subfigure}[b]{0.47\linewidth} 
        \centering
        \includegraphics[width=\linewidth]{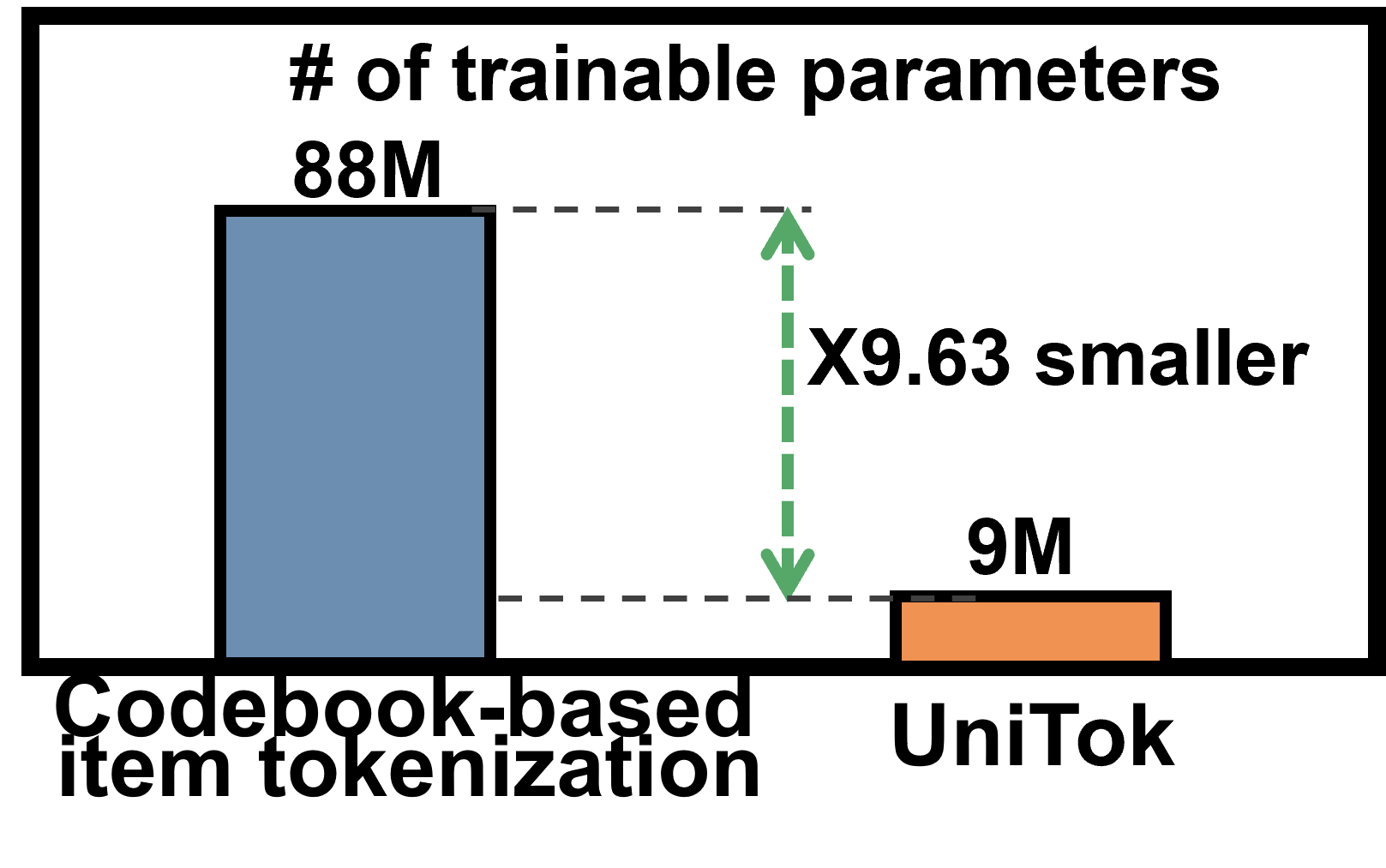} 
        \captionsetup{font={small,stretch=0.5}, skip=-5pt,textfont=normalfont}
        \caption{}
        \label{fig:c_a}
    \end{subfigure}
     \begin{subfigure}[b]{0.47\linewidth} 
        \centering
        \includegraphics[width=\linewidth]{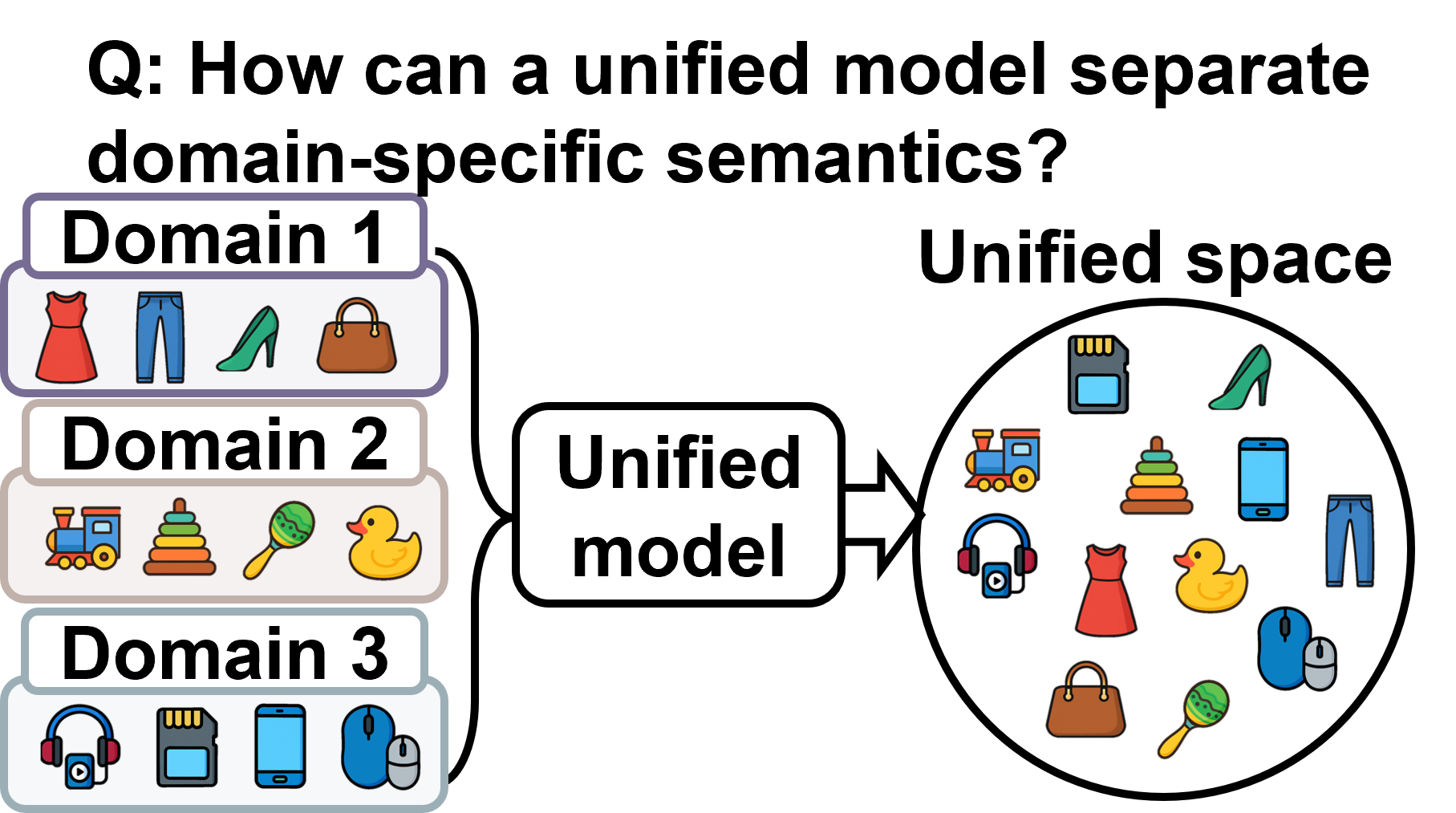} 
        \captionsetup{font={small,stretch=0.5}, skip=5pt,textfont=normalfont}
        \caption{}
        \label{fig:c_b}
    \end{subfigure}
    \caption{Examples illustrating (a) a comparison of the number of trainable parameters between codebook-based item tokenization methods and our method, \textsf{UniTok}, when applied to 10 distinct item domains, and (b) the inherent challenge of item tokenization across multiple domains.}
    \label{fig:challenges}
\end{figure}

Inspired by this, a natural question arising is: “How can we design a unified item tokenization framework for LLM-based recommendation that can be effectively generalized across multiple domains with minimal computational overhead?” To answer this question, we would like to outline the following two design challenges:
\begin{itemize}
    \item {\bf C1. Training overhead}: {\it Repeatedly training} domain-specific tokenizers is inefficient and resource-intensive. As shown in Figure \ref{fig:c_a}, when applied to 10 distinct domains, our \textsf{UniTok} method reduces the total number of trainable parameters by 9.63× compared to codebook-based item tokenization methods \cite{rajput2023recommender, wang2024learnable}, which require training a separate set of codebooks for each dataset to quantize items.
    \item {\bf C2. Semantic alignment}: The tokenizer must capture {\it rich semantics} from diverse domains; however, na\"ively using a shared token space across domains can cause semantic mixing and biased token assignments. Figure \ref{fig:c_b} exemplifies this challenge.
\end{itemize}
    
To address these aforementioned challenges, we make the first attempt towards developing a \underline{{\bf Uni}}fied item \underline{{\bf Tok}}enization framework designed to work effectively across multiple domains, named \textbf{\textsf{UniTok}}.

(\underline{{\bf Idea 1}}): Different domains often have exhibit distinct data distributions, requiring models to be trained separately to capture domain-specific patterns. To move beyond this limitation, we aim to design a unified item tokenization model capable of handling multiple domains without losing domain-specific knowledge. Achieving this goal requires the model to internally disentangle domain-specific learning from shared representations. To this end, we propose a new mixture-of-experts (MoE) architecture, dubbed TokenMoE, wherein domain-specific experts specialize in modeling patterns unique to each domain, while a shared expert captures the common knowledge across multiple domains. This architectural design enables the unified model to retain domain specialization without sacrificing global knowledge sharing (\underline{solving {\bf C1} and partially contributing to {\bf C2}}).

(\underline{{\bf Idea 2}}): To preserve item semantics during tokenization, we adopt a codebook-based approach \cite{rajput2023recommender} embedded within our MoE architecture, allowing each expert to specialize in distinct semantic patterns. However, the central challenge in multi-domain settings lies in ensuring semantic balance across diverse domains \cite{MaTZCLW0G22}. To address this, we introduce a mutual information (MI) calibration mechanism that explicitly encourages latent embeddings from each domain to retain sufficient and consistent informativeness. By minimizing the variance of MI across item domains, this mechanism attenuates inter-domain performance variability, enabling more stable and consistent generalization across diverse semantic spaces (\underline{solving {\bf C2}}).

Our main contributions are summarized as follows:
\begin{itemize}
    \item {\bf New methodology:} We propose \textsf{UniTok}, a unified item tokenization framework that integrates our customized MoE with codebooks to extract the semantic tokens for items across multiple domains while maintaining semantic balance.
    \item {\bf Extensive evaluations:} Through comprehensive experimental evaluations on diverse real-world datasets, we demonstrate (a) the superiority of \textsf{UniTok} in multi-domain scenarios, achieving substantial improvements of up to 51.89\% in NDCG@10, (b) the efficiency of \textsf{UniTok}, with a 9.63× reduction in model size compared to competitors, and (c) the strong generalization capability of \textsf{UniTok}, exhibiting robust performance in zero-shot settings without additional retraining.
    \item \textbf{Theoretical justifications:} We theoretically prove that \textsf{UniTok} (a) induces a higher entropy token space, (b) achieves a lower quantization error, and (c) ensures semantic consistency across domains by reducing the MI variance, fostering stable and balanced performance.
\end{itemize}

We refer readers to the supplementary materials for technical details omitted due to page limits.

\section{Preliminaries}

\subsection{Mixture-of-Experts}
The mixture-of-experts (MoE) architecture \cite{jacobs1991adaptive, shazeer2017outrageously, fedus2022switch, dai2024deepseekmoe} consists of multiple expert networks, each specializing in handling different parts of the input space. Formally, given an input $\mathbf{x}$, the output of an MoE model is a weighted combination of expert modules: $ \text{MoE}(\mathbf{x}) = \sum_{k=1}^{K} G_k(\mathbf{x}) E_k(\mathbf{x})$, where $K$ is the number of experts, $E_k(\mathbf{x})$ denotes the output of the $k$-th expert, and $G_k(\mathbf{x})$ is a softmax-based router function that assigns a probability to the $k$-th expert for given input. By dynamically routing input to specialized experts, MoE models can achieve greater model capacity and computational efficiency.

The MoE architecture is not only effective for model scaling but also well-suited for training on a large mixture of datasets \cite{jain2023damex}. Its gating mechanism enables adaptive expert selection for each dataset, allowing the model to generalize across diverse sources while minimizing interference between distributions.

\begin{figure*}[t]
    \centering
    \includegraphics[width=0.95\textwidth]{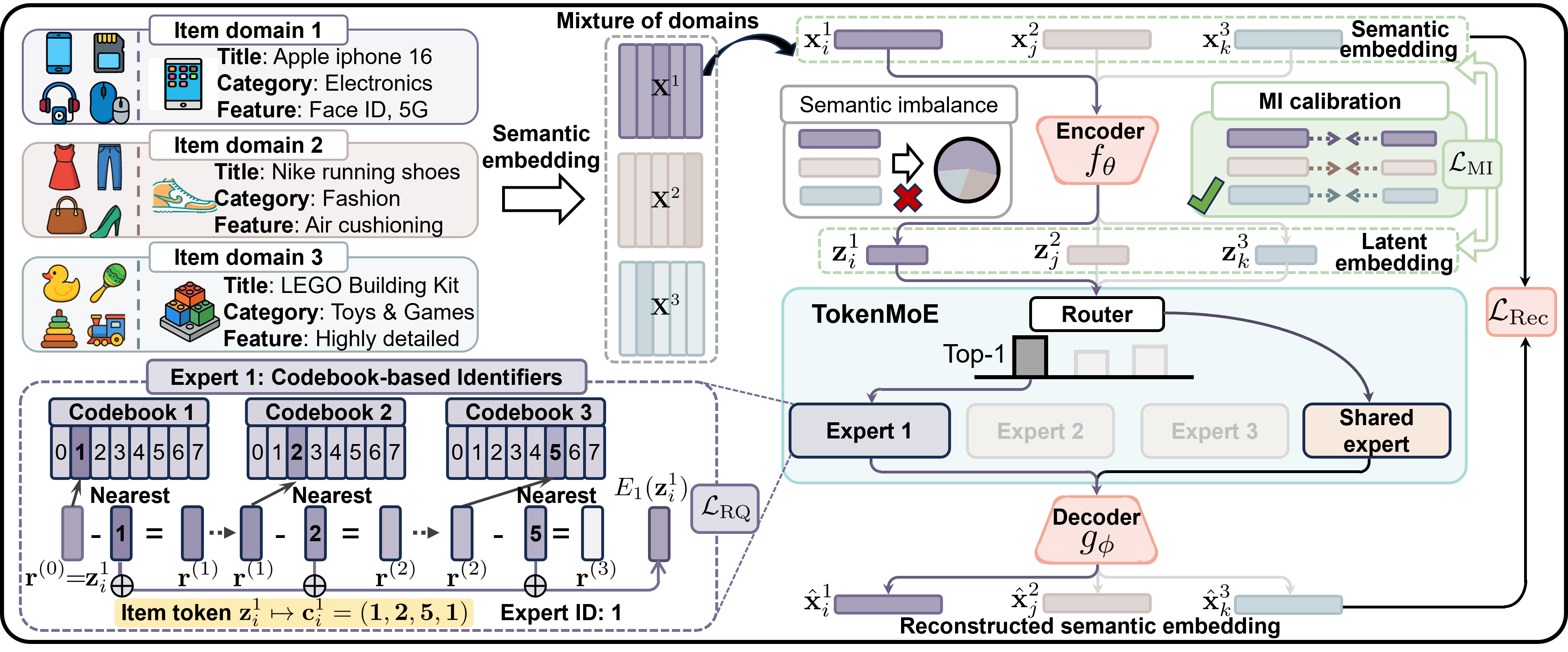}
    \caption{The schematic overview of the proposed \textsf{UniTok} framework.}
    \label{fig:overview}
\end{figure*}

\subsection{Codebook-based Identifiers}
Codebook-based identifiers \cite{rajput2023recommender} are generated using residual quantization (RQ), which encodes item metadata into hierarchical code sequences by sequentially applying codebooks to residuals across multiple stages. Given an input vector $\mathbf{x}$, the process starts with an initial residual $\mathbf{r}^{(0)} = \mathbf{x}$, and RQ recursively quantizes it using a series of $L$ codebooks $\{C_1, C_2, \dots, C_L\}$, where $C_\ell \triangleq \{\mathbf{c}_t\}_{t=1}^T$ contains $T$ code vectors at level $\ell$. The process is defined as
\begin{align}
    \mathbf{c}_\ell &= \arg\min_{\mathbf{c} \in C_\ell} \left\| \mathbf{r}^{(\ell-1)} - \mathbf{c} \right\|^2, \\
    \mathbf{r}^{(\ell)} &= \mathbf{r}^{(\ell-1)} - \mathbf{c}_\ell,
\end{align}
where $\mathbf{r}^{(\ell)}$ is the residual at level $l$, and $\mathbf{c}_\ell$ is the nearest code vector selected from codebook $C_\ell$. The final approximation of $\mathbf{x}$ after $L$ quantization stages is given by $\hat{\mathbf{x}} = \sum_{\ell=1}^{L} \mathbf{c}_{\ell}$.

Each selected code vector $\mathbf{c}_\ell$ corresponds to a discrete index $z_\ell \in \{1, \cdots, T\}$, resulting in a token sequence $(z_1, \cdots, z_L)$ that compactly represents the original input. This hierarchical quantization mechanism provides the foundation for our codebook-based identifiers, offering compact and semantically meaningful tokens well-suited for item tokenization. In recommender systems, such discrete tokens can then be directly used as input to LLMs~\cite{rajput2023recommender, wang2024learnable}, bridging the gap between item and language spaces while preserving semantic structure.

\section{Methodology}

\subsection{Task Formulation}

\subsubsection{Multi-domain setting.} Let $\mathcal{D}=\{\mathcal{D}_1,\dots,\mathcal{D}_K\}$ be a mixture of recommendation datasets for $K$ distinct item domains, where each domain $\mathcal{D}_k$ contains an item set $\mathcal{I}_k$ with associated textual metadata, such as titles, categories, features. Unlike typical single-domain scenarios, multi-domain settings pose significant challenges due to distributional inconsistencies \cite{jain2023damex}, making it crucial to handle such variability in item tokenization. Addressing this issue is essential for developing a scalable and unified item tokenization method that supports multiple domains, particularly in LLM-based recommender systems. Although collaborative signals can enhance recommendation performance, they inherently rely on user--item interactions, introducing computational overhead and limiting generalization when {\it shared users are absent} across domains. Instead, we aim to develop a unified tokenization method that operates independently of user data, enabling scalable and general-purpose recommendations in multi-domain LLM-based systems.

\noindent
\subsubsection{Item tokenization task.} Given raw items $\mathcal{I}_k$ from any domain-specific dataset $\mathcal{D}_k$, we assume that we have access to a pre-trained content encoder to generate the semantic embeddings for domain $k$, denoted as $\mathbf{X}^k \in \mathbb{R}^{|\mathcal{I}_k| \times d}$, where $|\mathcal{I}_k|$ is the number of items in domain $k$ and $d$ is the embedding dimensionality. The $i$-th embedded item is represented as $\mathbf{x}_i^k \in \mathbf{X}^k$. The objective is to learn a mapping function $\mathcal{F}: \mathbb{R}^d \rightarrow \mathcal{C}$ that projects each continuous embedding $\mathbf{x}_i^k$ into a discrete codeword $\mathbf{c}_i^k \in \mathcal{C}$, where $\mathcal{C}$ denotes the shared space of discrete item tokens across all domains.

\subsection{Overview of \textsf{UniTok}}

We recall that recent item tokenization methods for LLM-based recommendations focus merely on a single-domain setting \cite{rajput2023recommender, wang2024learnable, zheng2024adapting}. Howerer, real-world systems recently operate across multiple domains \cite{jiang2022adaptive, ning2023multi}, leading to repeated training and semantic inconsistency across domains---an issue largely overlooked by prior studies.

To tackle challenges \textbf{C1} and \textbf{C2} in Section 1, \textsf{UniTok} leverages a shared autoencoder to project items in the mixture of domains into a unified latent space. To achieve effective tokenization across diverse domains, we present TokenMoE with codebook-based identifiers, an MoE architecture composed of domain-specific experts that capture specialized semantics and a shared expert that encodes generalized knowledge. Additionally, we present an MI-based loss that enforces consistent semantics across multiple domains by regulating the informativeness of latent embeddings.

\subsection{Architectural Details}

As illustrated in Figure \ref{fig:overview}, our \textsf{UniTok} consists of four key components: a shared autoencoder, TokenMoE, codebook-based identifiers, and an MI calibration mechanism. 

\subsubsection{Shared autoencoder.}
Given semantic embeddings from the mixture of domains, we first employ a shared autoencoder composed of an encoder $f_\theta$, to project items into a unified latent space, and decoder $g_\phi$ to reconstruct the semantic embeddings (see the coral pink region in Figure \ref{fig:overview}). This establishes a unified representation space across multiple domains, which captures common structural patterns while retaining essential information for reconstruction.

Formally, for each input item $\mathbf{x}_i^k \in \mathbf{X}^k$ from domain $\mathcal{D}_k$, the encoder produces a latent embedding $\mathbf{z}_i^k = f_{\theta}(\mathbf{x}_i^k)$, and the decoder reconstructs the input item as $\hat{\mathbf{x}}_i^k = g_\phi(\hat{\mathbf{z}}_i^k)$, where $\hat{\mathbf{z}}_i^k = \text{TokenMoE}\left( \mathbf{z}_i^k \right)$ (to be specified in Eq. (\ref{eq:tokenmoe})). The model is optimized using the reconstruction loss: 
\begin{equation}
    \mathcal{L}_\text{Rec} = \sum\limits_{k = 1}^K {\sum\limits_{\mathbf{x}_i^k  \in \mathbf{X}^k } {\left\| {\mathbf{x}_i^k  - \hat{\mathbf{x}}_i^k } \right\|^2 } } ,
\end{equation}
where $\left\| \cdot \right\|$ denotes the $L_2$ norm. This reconstruction loss shapes the latent space into a compact and informative representation that retains core semantics for item tokenization.

\subsubsection{TokenMoE.}
Conventional tokenization models applied to a unified item space often fail to capture domain-specific nuances, as they treat diverse domains uniformly, potentially leading to the loss of specialized semantic information. To address this limitation, we present TokenMoE, a more generalizable MoE architecture, which routes items to both domain-specific experts and a shared expert. This design enables domain-specific experts to learn specialized patterns, while a shared expert with always-active routing facilitates efficient knowledge transfer across multiple domains. Distinct from earlier approaches that utilize MoE within the feedforward layers of transformers \cite{dai2024deepseekmoe}, our key contribution lies in uniquely integrating MoE into the tokenization module to better enable domain-aware tokenization. The TokenMoE module is illustrated in the light blue region of Figure \ref{fig:overview} (with one of the domain-specific expert highlighted in purple and the shared expert in orange). This design addresses {\bf C1} and partially mitigates {\bf C2}.

Specifically, after encoding, the item latent embedding $\mathbf{z}_i^k$ passes through a router function $G(\cdot)$, which produces a softmax distribution over $K$ domain-specific experts: $    G(\mathbf{z}_i^k) \!=\! \{G_1,\! G_2, \ldots,\! G_K\}$, where each $G_k$ is computed as 
\begin{equation}
\begin{aligned}
    G_k &= \frac{\exp(s_i^{(k)})}{\sum_{j=1}^K \exp(s_i^{(j)})}, s_i = h(\mathbf{z}_i^k) \in \mathbb{R}^K,
\end{aligned}
\end{equation}
and $h(\cdot)$ is a learnable linear transformation in the router producing the router logits $s_i$, and $s_i^{(k)}$ denotes the logit corresponding to the $k$-th domain-specific expert. Here, $K$ is the total number of domain-specific experts, which is typically aligned with the number of domains.

The item is then routed to the top-$N$ domain-specific experts\footnote{$N$ is typically set as either 1 or 2  to promote sparsity in expert activation, which significantly reduces computational overhead while preserving model capacity \cite{lepikhin2020gshard, fedus2022switch}.} based on the highest values of $G(\mathbf{z}_i^k)$, while it is also deterministically assigned to a shared expert. Each expert, including the shared one, is implemented as a codebook-based identifier (to be specified later). The final $\hat{\mathbf{z}}_i^k$ is computed as a weighted combination of the selected domain-specific experts and the shared expert, which is formulated as follows:
\begin{equation}
\left\{ \begin{array}{l}
 {\hat{\mathbf{z}}}_i^k  = \text{TokenMoE} \left( \mathbf{z}_i^k\right) = \sum\limits_{k = 1}^K {G_k } E_k ({\mathbf{z}}_i^k ) + E_{{\text{share}}} ({\mathbf{z}}_i^k ), \\ 
 G_k  = \left\{ {\begin{array}{*{20}c}
   {G_k ,} \hfill & {{\text {if }} k \in {\text{Top}}_N (G({\mathbf{z}}_i^k ))} \hfill  \\
   {0,} \hfill & {{\text{otherwise}}} \hfill  \\
\end{array}} \right., \\ 
 \end{array} \right.
 \label{eq:tokenmoe}
\end{equation}
where, $E_k(\cdot)$ denotes the $k$-th expert module, $E_\text{share}(\cdot)$ denotes the shared expert module, and $\text{Top}_N\left( {G \left( \mathbf{z}_i^k \right)} \right) = \left\{ {k_1 ,k_2 , \ldots ,k_N } \right\} $ is the set of indices corresponding to the top-$N$ experts selected by the router. Figure \ref{fig:overview} shows an example where the top-1 expert is selected.

TokenMoE routes each item to domain-specific experts via its learned router, while a shared expert captures common knowledge through a deterministic path. To encourage expert specialization, each expert is initialized with the mean feature of a specific domain \cite{wang2024learnable}, providing a strong inductive bias for domain-aware tokenization {\it without requiring explicit supervision}. By activating only a subset of experts per item, TokenMoE enhances scalability, maintains domain specificity, and supports better generalization.

\subsubsection{Codebook-based identifiers.}
 We adopt RQ \cite{rajput2023recommender} in each expert to discretize each item into compact token sequences. Given an item latent embedding $\mathbf{z}_i^k \in \mathbb{R}^d$ produced by the shared encoder, RQ approximates it through a sequence of codebooks $\{C_1, C_2, \dots, C_L\}$, where $L$ is the codebook size, and each codebook $C_\ell \triangleq \{\mathbf{c}_t\}_{t=1}^T$ contains $T$ code vectors. As shown in the bottom-left of Figure \ref{fig:overview}, at each level $\ell$, the residual $\mathbf{r}^{(\ell)}$ from the previous step is encoded using the nearest code $\mathbf{c}_\ell$ in $C_\ell$. The sum of all selected codes reconstructs the original latent embedding:
\begin{equation}
    E_k(\mathbf{z}_i^k) \approx \sum\nolimits_{\ell=1}^L \mathbf{c}_\ell, \quad \text{where } \mathbf{c}_\ell \in C_\ell.
\end{equation}
This hierarchical quantization enables fine-grained and memory-efficient tokenization, as each item is tokenized by a discrete codeword:
\begin{equation}
    \mathbf{z}_i^k \mapsto \mathbf{c}_i^k = (z_1, \dots, z_L, e_1, ...,e_N),
\end{equation}
where $z_\ell \in \{1, \dots, T\}$ denotes the index of selected code vector $\mathbf{c}_\ell$ from the $\ell$-th codebook $C_\ell$ and $e_n \in \{1, \dots, K\}$ indicates the expert ID of the $n$-th top-$N$ expert chosen by the router.

To train this quantization process, we adopt the RQ loss:
\begin{equation}
        \mathcal{L}_{\text{RQ}} := \sum\nolimits_{\ell=1}^{L} \left\| \text{sg}[\textbf{r}^{(\ell)}] - \mathbf{c}_\ell \right\|^2 + \alpha \left\| \mathbf{r}^{(\ell)} - \text{sg}[{\mathbf{c}_\ell}] \right\|^2,
        \label{eq: l_rq}
\end{equation}
where $\mathbf{r}^{(\ell)}$ is the residual vector at level $\ell$, ${\mathbf{c}_\ell}$ is the selected code vector from $C_\ell$, $\text{sg}[\cdot]$ is the stop-gradient operator, and $\alpha$ is a balancing hyperparameter. In Eq. (\ref{eq: l_rq}), the first term aligns the code vector to the target residual (codebook learning), and the second term forces the encoder and router to commit to the selected quantized code vector.

\subsubsection{MI calibration.}
As the shared encoder will project all items into a unified latent embedding space, it is not straightforward for the model to precisely capture domain-specific features due to semantic imbalance. This occurs when the quality of learned latent embeddings varies significantly across diverse domains—particularly between simple and complex domains—causing semantically similar items to be assigned inconsistent tokens \cite{MaTZCLW0G22}.

\begin{figure}[t]
    \centering
    \includegraphics[width=0.99\columnwidth]{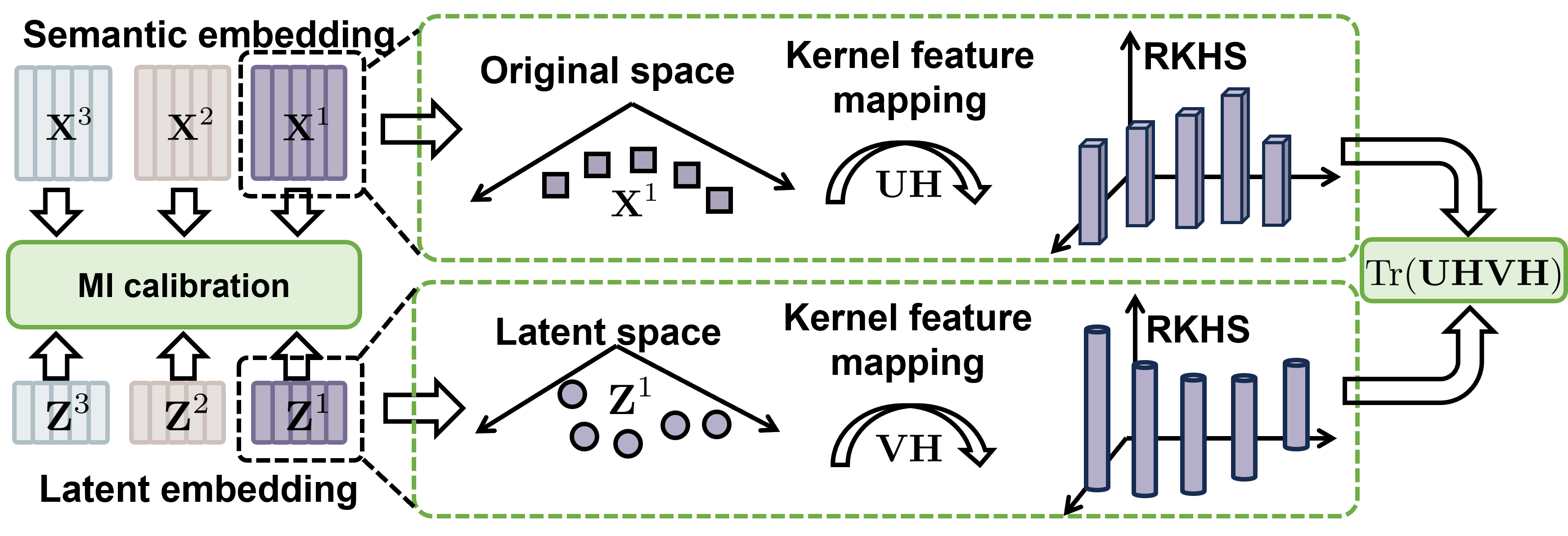}
    \caption{The illustration of MI calibration.}
    \label{fig:MI}
\end{figure}

As another key contribution aimed at mitigating this issue, we introduce an MI mechanism to ensure that the latent space retains sufficient information from each domain. Specifically, we adopt the Hilbert-Schmidt independence criterion (HSIC)~\cite{gretton2005measuring, li2021self}\footnote{Other methods, such as MINE~\cite{belghazi2018mutual} and InfoNCE \cite{oord2018representation}, can also estimate MI using neural networks, but require additional training. To maintain simplicity, we adopt HSIC, which is non-parametric.}, serving as a proxy for the MI, with a higher HSIC value indicates stronger dependence. As illustrated in Figure \ref{fig:MI}, for domain $k$, HSIC measures the dependence between the input semantic embeddings $\mathbf{X}^k = \{\mathbf{x}_1^k, \dots, \mathbf{x}_{|\mathcal{I}_k|}^k\}$ and their latent embeddings $\mathbf{Z}^k = \{\mathbf{z}_1^k, \dots, \mathbf{z}_{|\mathcal{I}_k|}^k\}$ in a reproducing kernel Hilbert space (RKHS), computed as:
\begin{equation}
    \widehat{\operatorname{HSIC}}(\mathbf{X}^k, \mathbf{Z}^k) = \frac{1}{(|\mathcal{I}_k| - 1)^2} \operatorname{Tr}(\mathbf{U} \mathbf{H} \mathbf{V} \mathbf{H}),
\end{equation}
where $\mathbf{U}, \mathbf{V} \in \mathbb{R}^{|\mathcal{I}_k| \times |\mathcal{I}_k|}$ are Gaussian kernel matrices computed over $\mathbf{X}^k$ and $\mathbf{Z^k}$, respectively\footnote{The kernel matrices are computed element-wise using the Gaussian kernel: $U_{ij} \!\! = \!\!u(\mathbf{x}_i^k ,\mathbf{x}_j^k ) \!\!= \!\!\exp \left( {{{ \!- \|\mathbf{x}_i^k  \!-\! \mathbf{x}_j^k\| ^2 }\! / {2\sigma ^2\! }}} \right)$ and $V_{ij} \!\! = \!\!v(\mathbf{z}_i^k ,\mathbf{z}_j^k ) \!\!= \!\!\exp \!\left( { \!- {{\|\mathbf{z}_i^k \! - \!\mathbf{z}_j^k\| ^2 } / {2\sigma ^2 }}}\right)$ for the bandwidth $\sigma$.}. The centering matrix $\mathbf{H} = \mathbf{I} - \frac{1}{|\mathcal{I}_k|} \mathbf{1}\mathbf{1}^\top$ ensures zero-mean embeddings in RKHS; and $\operatorname{Tr}(\mathbf{UHVH})$ computes the Hilbert-Schmidt norm of the cross-covariance between the two RKHSs. This design addresses {\bf C2}.

To enforce semantic balance across multiple domains (see Figure \ref{fig:MI}), we characterize the MI calibration loss as:
\begin{equation}
    \mathcal{L}_{\text{MI}} = \operatorname{Var}\left[ \widehat{I}^{(k)} \right] - \beta \mathbb{E}\left[ \widehat{I}^{(k)} \right],
    \label{eq: l_mi_calib}
\end{equation}
where $\widehat{I}^{(k)} = \widehat{\mathrm{HSIC}}(\mathbf{X}^k, \mathbf{Z}^k)$ and $\beta$ is a weighting hyperparameter. The first term penalizes high variance of MI across domains to mitigate semantic imbalance, while the second term enforces each domain to retain sufficient domain-specific information. 

\subsubsection{Optimization.} We train the model using the overall loss:
\begin{equation}
    \mathcal{L}_\text{total} = \mathcal{L}_\text{Rec} +  \lambda_\text{RQ}\mathcal{L}_\text{RQ} + \lambda_\text{MI} \mathcal{L}_\text{MI},
    \label{eq:overall_loss}
\end{equation}
where $ \lambda_\text{RQ}$ and $ \lambda_\text{MI}$ are hyper-parameters to control the strength of RQ and MI, respectively.

To instantiate \textsf{UniTok} in LLM-based recommender systems, we first train \textsf{UniTok} on all items using Eq. (\ref{eq:overall_loss}). The trained \textsf{UniTok} then tokenizes each item into discrete semantic tokens, enabling LLMs to operate in the token space. Following prior work \cite{wang2024learnable}, user interaction histories $\mathbf{u}$ are converted into item token sequences $\mathbf{u} = [\tilde{\mathbf{c}}_1, \tilde{\mathbf{c}}_2, \dots, \tilde{\mathbf{c}}_P]$, and the LLM-based recommender system learns to predict the next interacted item token $\tilde{\mathbf{c}}_{P+1}$.

\subsection{Theoretical Analyses}

In this subsection, we establish the following theorems, which analyze and support the effectiveness of the key components within \textsf{UniTok}. 

\textbf{Theorem 1.} \textit{The token space induced by \textsf{UniTok} exhibits strictly higher entropy than that of standard codebook-based methods:}
\begin{equation}
    H(\mathcal{C}_{\textsf{UniTok}}) > H(\mathcal{C}_{\text{standard}}),
\end{equation}
\textit{where $\mathcal{C}_{\text{UniTok}}$ and $\mathcal{C}_{\text{standard}}$ denote the discrete token distributions generated by \textsf{UniTok} and standard codebook-based methods, respectively.}

This indicates that judiciously incorporating multiple experts increases the overall entropy, thereby expanding the token space capacity.

\textbf{Theorem 2.} \textit{Let $\mathbb{E}[\mathcal{L}_\textsf{UniTok}]$ and $\mathbb{E}[\mathcal{L}_{\text{standard}}]$ denote the expected quantization error of \textsf{UniTok} and standard codebook-based methods using a single shared codebook, respectively. Then, the following inequality holds:}
\begin{equation}
    \mathbb{E}[\mathcal{L}_{\textsf{UniTok}}] \leq \mathbb{E}[\mathcal{L}_{\text{standard}}].
\end{equation}

This implies that a lower expected quantization error reflects more precise modeling of item tokenization. The TokenMoE framework further reduces this error by leveraging expert specialization, which effectively compensates for domain-specific inaccuracies to some extent, thereby improving quantization quality across diverse domains.

\textbf{Theorem 3.}
\textit{Suppose that the loss $\mathcal{L}^{(k)}$ on the $k$-th domain is Lipschitz-continuous with respect to the informativeness of representations. Then, the performance variability across domains is upper-bounded by the variance of MI:}
\begin{equation}
    |\mathcal{L}^{(i)} - \mathcal{L}^{(j)}| \leq C \sqrt { \text{Var}\left[ {\widehat{I}^{(k)}} \right]}, \forall i, j
\end{equation}
\textit{where $\text{Var}\left[ {\widehat{I}^{(k)}} \right]$ is the variance of MI estimates across domains and $C$ is a constant.}

This implies that reducing MI variance across domains promotes consistent semantic representations, leading to more stable and reliable downstream performance.


We empirically validate each theorem’s technical correctness and practical relevance through extensive experiments.

\begin{table*}[!t]\centering
\setlength\tabcolsep{6.0pt}
\small
  \begin{tabular}{c|c|c|c|c|c|c|c|c|c|c}
    \toprule[1pt]
    \multicolumn{1}{c|}{Method}&\multicolumn{1}{|c|}{{\bf Beauty}}&\multicolumn{1}{c|}{{\bf Cellphones}}&\multicolumn{1}{c}{{\bf Grocery}}&\multicolumn{1}{|c|}{{\bf Instruments}}&\multicolumn{1}{|c|}{{\bf Office}} & {\bf Pet Supplies} & {\bf Tools} & {\bf Toys} & {\bf Games} & {\bf Yelp}\\
    \midrule[1pt]
    {\bf MF} & 0.0369 & 0.0267 & 0.0216 & 0.0710 & 0.0255 & 0.0268 & 0.0169 & 0.0192 & 0.0366 & 0.0144\\
    {\bf LightGCN} & 0.0285 & 0.0456 & 0.0357 & 0.0781 & 0.0301 & 0.0289 & 0.0257 & 0.0287 & 0.0417 &0.0195\\
    {\bf SASRec} & 0.0314 & 0.0446 & 0.0376 & 0.0609 & 0.0285 & 0.0301 & 0.0234 & 0.0239 & 0.0412 & 0.0183\\
    {\bf Bert4Rec} & 0.0194 & 0.0268 & 0.0237 & 0.0573 & 0.0274 & 0.0161 & 0.0092 & 0.0177 & 0.0379 & 0.0131 \\
    \midrule
    {\bf P5-TID} & 0.0255 & 0.0357 & 0.0316 & 0.0721 & 0.0239 & 0.0243 & 0.0198 & 0.0202 & 0.0388 & 0.0154\\
    {\bf P5-SemID} & 0.0304 & 0.0406 & 0.0351 & 0.0730 & 0.0283 & 0.0282 & 0.0237 & 0.0231 & 0.0432 & 0.0188\\
    {\bf TIGER} & 0.0324 & 0.0446 & 0.0375 & 0.0788 & 0.0295 & 0.0279  & 0.0284 & 0.0268  & 0.0427  & 0.0208\\
    {\bf LC-Rec} & \underline{0.0381}& 0.0458& 0.0369& 0.0802& 0.0311 & \underline{0.0335} & \underline{0.0307}& 0.0279& 0.0451& 0.0215\\
    {\bf LETTER} & 0.0364& \underline{0.0473}& \underline{0.0392} & \underline{0.0831} & \underline{0.0326} & 0.0307 & 0.0298 & \underline{0.0291} & \underline{0.0469} & \underline{0.0231} \\
    \rowcolor{lightgray!30}
    {\bf \textsf{UniTok}}&  {\bf 0.0478}& {\bf 0.0647}& {\bf 0.0533}& {\bf 0.0884} & {\bf 0.0432}  & {\bf 0.0496} & {\bf 0.0439} & {\bf 0.0442} & {\bf 0.0476} & {\bf 0.0321}\\
    \midrule
    {\bf Gain}& 25.46\% & 36.78\% & 35.97\% & 6.38\% & 32.52\%  & 48.06\% & 42.99\% & 51.89\% & 1.49\% & 38.96\% \\
    \bottomrule[1pt]
  \end{tabular}
    \caption{Performance comparison among \textsf{UniTok} and recommendation competitors for the ten benchmark datasets in terms of NDCG@10. Here, the best and second-best performers are highlighted by bold and underline, respectively. The improvements are statistically significant on average ($p = 0.0219 < 0.05$) based on paired t-tests over five runs across all datasets.}
\label{tab:RQ1}
\end{table*}

\section{Experimental Evaluation}

In this section, we carry out comprehensive experiments to empirically validate the effectiveness of \textsf{UniTok} across multiple domains.

\subsection{Experimental Settings}

\subsubsection{Datasets.} We conduct our experiments on ten real-world datasets spanning ten domains widely adopted for evaluating the performance of recommendations, which include Beauty, Cellphones, Grocery, Instruments, Office, Pet Supplies, Tools, Toys, Games\footnote{https://nijianmo.github.io/amazon/index.html.}, and Yelp\footnote{https://www.yelp.com/dataset.}. Each item contains metadata, including title, category, and features. Due to page limitations, we report complete results across all datasets to assess overall performance, while employing three representative datasets for the efficiency and ablation analyses.

\subsubsection{Competitors.} To comprehensively demonstrate the superiority of \textsf{UniTok}, we present nine benchmark recommendation methods, including four widely-used collaborative filtering methods, MF~\cite{rendle2012bpr}, SASRec~\cite{kang2018self}, LightGCN~\cite{he2020lightgcn}, Bert4Rec~\cite{sun2019bert4rec}, and five item tokenization-aided recommendations, P5-TID~\cite{hua2023index}, P5-SemID~\cite{hua2023index}, TIGER~\cite{rajput2023recommender}, LC-Rec~\cite{zheng2024adapting}, and LETTER~\cite{wang2024learnable}. 

\subsubsection{Performance metrics.} Following the full-ranking protocol~\cite{he2020lightgcn}, we rank all non-interacted items for each user and evaluate using Recall@$M$ (R@$M$) and NDCG@$M$ (N@$M$), where $M \in \left\{ {5,10} \right\} $.

\subsubsection{Implementation details.}

For item tokenization, we use 4-level codebook-based identifiers, where each codebook comprises 256 code vectors with a dimension of 32. We set $\lambda_\text{RQ}$ to 1 and $\lambda_\text{MI}$ to 0.03. All experiments are conducted on two NVIDIA RTX 3090 GPUs. 

\subsection{Can One Tokenizer Serve All Domains?}

To evaluate the recommendation accuracy of \textsf{UniTok}, we compare it with benchmark item tokenization methods across ten benchmark datasets from diverse domains. Notably, distinct from benchmark methods that train separate tokenizers for each dataset, \textsf{UniTok} trains a {\it single unified model} that handles all ten datasets jointly. As shown in Table~\ref{tab:RQ1}, \textsf{UniTok} consistently outperforms competitors across all datasets, validating the effectiveness of our unified tokenization framework, achieving up to 51.89\% improvement in terms of NDCG@10 on the Tools dataset. Unlike other approaches that require domain-specific customization, \textsf{UniTok} effectively captures item semantics across diverse domains through a single, shared tokenization process—highlighting the strength and generality of our proposed design.

We observe that item tokenization-aided LLM-based recommender systems, including TIGER, LC-Rec, LETTER, and \textsf{UniTok}, are superior to standard collaborative filtering methods such as MF, LightGCN, SASRec, and Bert4Rec. This improvement benefits from the rich semantic understanding and reasoning capabilities inherent in LLMs, which go beyond conventional user--item interactions. Moreover, integrating carefully designed item tokenization into LLM-based recommender systems yields additional performance gains compared to LLMs that solely on item metadata for tokenization ({\it e.g.}, P5-TID and P5-SemID). This is because item tokenization serves as a critical bridge between the item space and the language space, allowing the underlying model to represent items in a discrete, language-aligned semantic space that enhances generalization and reasoning.

\subsection{Is \textsf{UniTok} More Efficient than Traditional Tokenizers?}

\begin{table}[t]
\small
    \centering
    \begin{tabular}{l|cc}
    \toprule
    \textbf{Module} & \textbf{Codebook-based methods} & \textbf{\textsf{UniTok}} \\
    \midrule
    \textbf{Codebook} & 0.33M & 0.36M\\
    \textbf{Autoencoder} & 87.45M & 8.75M\\
    \textbf{Router} & -- & 0.01M\\
    \textbf{Total} & 87.78M & \cellcolor{lightgray} 9.11M\\
    \bottomrule
    \end{tabular}
    \caption{Comparison of the number of trainable parameters. For competitors, we report the total number of trainable parameters accumulated across all ten datasets.}
    \label{tab:param_comparison}
\end{table}

To validate the efficiency of \textsf{UniTok}, we compare the total number of trainable parameters between \textsf{UniTok} and traditional codebook-based competitors, including TIGER, LC-Rec, and LETTER, which share the same underlying architecture. While these competitors require training and storing separate tokenization models for each dataset, \textsf{UniTok} employs a single unified model shared across all datasets. For a fair comparison, we report the cumulative trainable parameter count of the codebook-based competitors over all datasets. As shown in Table~\ref{tab:param_comparison}, \textsf{UniTok} achieves approximately a tenfold reduction in the total number of trainable parameters. This efficiency comes primarily from using the shared autoencoder, while the number of additional trainable parameters introduced by the codebook and TokenMoE router remains negligible. By eliminating the need for domain-specific tokenization learning, \textsf{UniTok} offers substantial advantages in scalability and deployment efficiency.

We further evaluate the performance of competitors under a single unified training setup, where each competitor is trained jointly across ten datasets using a comparable number of trainable parameters to that of \textsf{UniTok}. As shown in Table~\ref{tab:baseline_unified}, the competing methods exhibit substantial performance degradation in this setting compared to their single-domain scenario (see Table \ref{tab:RQ1}), primarily due to the difficulty of distinguishing items from different domains when using shared tokenization. In contrast, \textsf{UniTok} maintains consistently superior recommendation performance, achieving up to 84.51\% improvement in Recall@10 on Cellphones, while using a {\it similar trainable parameter budget}. This efficiency stems from its modular TokenMoE architecture, where domain-specific experts learn semantics independently, while sharing a unified token space.

\begin{table}[!t]\centering
\small
\setlength\tabcolsep{2.0pt}
  \begin{tabular}{c|cc|cc|cc}
    \toprule[1pt]
    \multicolumn{1}{c|}{}&\multicolumn{2}{|c|}{{\bf Beauty}}&\multicolumn{2}{c|}{{\bf Cellphones
}}&\multicolumn{2}{c}{{\bf Grocery}}\\
    \cmidrule{1-7}
           {\bf Method} & {\bf R@10}& {\bf N@10}& {\bf R@10}& {\bf N@10}& {\bf R@10}& {\bf N@10}\\
    \midrule[1pt]
    {\bf TIGER} & 0.0499 & 0.0267 & 0.0661&0.0342 & 0.0576 & 0.0273\\
    {\bf LC-Rec} & \underline{0.0564}& \underline{0.0302} & 0.0647&0.0337& 0.0584 & 0.0287\\
    {\bf LETTER}& 0.0528& 0.0288& \underline{0.0678}& \underline{0.0363}& \underline{0.0618}& \underline{0.0315}\\
    \rowcolor{lightgray!30}
    {\bf \textsf{UniTok}}& {\bf 0.0934}& {\bf 0.0478}& {\bf 0.1251}& {\bf 0.0647}& {\bf 0.1061}& {\bf 0.0533}\\
    \midrule
    {\bf Gain}& 65.60\% & 58.28\% & 84.51\% & 78.23\% & 71.68\% & 69.21\% \\
    \bottomrule[1pt]
  \end{tabular}
    \caption{Performance comparison of \textsf{UniTok} and tokenization competitors under a single unified training setup. Here, the best and second-best performers are highlighted by bold and underline, respectively.}
  \label{tab:baseline_unified}
\end{table}

\subsection{Can \textsf{UniTok} be Generalized to Unseen Domains?}

To evaluate the generalization ability of \textsf{UniTok}, we adopt a {\it zero-shot} setting where our item tokenizer model, \textsf{UniTok}, is trained once on the ten source datasets and is directly tested on unseen datasets from three target domains—Clothing, Health, and Sports—without any additional training or fine-tuning. This setup reflects real-world scenarios where new domains may appear dynamically after the model has been deployed.

As shown in Table~\ref{tab:unseen}, \textsf{UniTok} significantly outperforms existing item tokenization-based recommender systems, achieving up to 17.87\% improvement in NDCG@10 on Health. While the competitors require retraining on each new dataset to achieve reasonable tokenization results, \textsf{UniTok} maintains robust accuracy without any further adaptation. This demonstrates our model’s ability to learn a discrete token space that captures transferable item semantics across diverse domains. The consistently strong zero-shot performance further highlights the robustness and practical utility of our unified tokenization framework in supporting effective generalization across heterogeneous domains.

\begin{table}[!t]\centering
\setlength\tabcolsep{2.0pt}
\small
  \begin{tabular}{c|cc|cc|cc}
    \toprule[1pt]
    \multicolumn{1}{c|}{}&\multicolumn{2}{|c|}{{\bf Clothing}}&\multicolumn{2}{c|}{{\bf Health
}}&\multicolumn{2}{c}{{\bf Sports}}\\
    \cmidrule{1-7}
           {\bf Method} & {\bf R@10}& {\bf N@10}& {\bf R@10}& {\bf N@10}& {\bf R@10}& {\bf N@10}\\
    \midrule[1pt] 
    {\bf TIGER} & 0.0501 & 0.0242 & 0.0677& 0.0342& 0.0469 & 0.0228\\
    {\bf LC-Rec} & \underline{0.0527}& \underline{0.0266}& 0.0694& 0.0358& 0.0494 & 0.0246\\
    {\bf LETTER}& 0.0515& 0.0257& \underline{0.0717}& \underline{0.0375}& \underline{0.0510}& \underline{0.0265}\\
    \rowcolor{lightgray!30}
    {\bf \textsf{UniTok}}& {\bf 0.0592}& {\bf 0.0288}& {\bf 0.0835}& {\bf 0.0442}& {\bf 0.0591}& {\bf 0.0298}\\
    \midrule
    {\bf Gain}& 12.33\% & 8.27\% & 16.46\% & 17.87\% & 15.88\% & 12.45\% \\
    \bottomrule[1pt]
  \end{tabular}
    \caption{Performance comparison among \textsf{UniTok} and recommendation competitors for the three unseen datasets. Here, the best and second-best performers are highlighted by bold and underline, respectively.}
  \label{tab:unseen}
\end{table}

\begin{table}[!t]\centering
\setlength\tabcolsep{3.3pt}
\small
  \begin{tabular}{c|cc|cc|cc}
    \toprule[1pt]
    \multicolumn{1}{c|}{}&\multicolumn{2}{|c|}{{\bf Beauty}}&\multicolumn{2}{c|}{{\bf Cellphones
}}&\multicolumn{2}{c}{{\bf Grocery}}\\
    \cmidrule{1-7}
           {\bf Method} & {\bf R@10}& {\bf N@10}& {\bf R@10}& {\bf N@10}& {\bf R@10}& {\bf N@10}\\
    \midrule[1pt]
    {\bf \textsf{UniTok-1}}& 0.0558 & 0.0304 &  0.0702& 0.0371 & 0.0633 & 0.0342\\
    {\bf \textsf{UniTok-2}}& 0.0896 & 0.0436 & 0.1194 & 0.0606 & 0.0989 & 0.0497 \\
    {\bf \textsf{UniTok-3}}& 0.0915& 0.0457& 0.1225& 0.0622& 0.1044& 0.0515\\
    \rowcolor{lightgray!30}
    {\bf \textsf{UniTok}}& {\bf 0.0934}& {\bf 0.0478}& {\bf 0.1251}& {\bf 0.0647}& {\bf 0.1061}& {\bf 0.0533}\\
    \bottomrule[1pt]
  \end{tabular}
    \caption{Ablation study results on the Beauty, Cellphones, and Grocery datasets.}
  \label{tab:ablation_main}
\end{table}

\subsection{What Makes \textsf{UniTok} Effective?}

To assess the contribution of each component in \textsf{UniTok}, we perform an ablation study by progressively removing or modifying its core modules: the TokenMoE module, the shared expert, and the MI calibration part. \textsf{UniTok-1} removes the TokenMoE and MI calibration, using only a single set of codebooks without any MoE structure; \textsf{UniTok-2} keeps TokenMoE, but removes the shared expert and MI calibration; and \textsf{UniTok-3} only removes the MI calibration. \textsf{UniTok} includes all components. As shown in Table~\ref{tab:ablation_main}, removing any module leads to a noticeable drop in recommendation accuracy, which confirms that the combination of modules is crucial to \textsf{UniTok}’s effectiveness. In particular, comparing \textsf{UniTok-1} and \textsf{UniTok-2} highlights the importance of the TokenMoE module, which significantly improves performance across all datasets by capturing domain-specific semantics. Additionally, in comparison with \textsf{UniTok-3}, introducing MI calibration further enhances performance by enforcing the learned latent embeddings to retain essential semantic information from each domain.

\section{Conclusions and Outlook}

We explored an open yet fundamental challenge in multi-domain LLM-based recommendations by building a unified item tokenization framework. To this end, we proposed \textsf{UniTok}, which integrates a customized MoE architecture with codebooks to generate semantically meaningful tokens across diverse domains. Experiments on wide-ranging datasets showed that \textsf{UniTok} (a) achieves up to 51.89\% gains in NDCG@10, (b) reduces trainable parameters by 9.63×, and (c) is theoretically validated with respect to the effectiveness of its individual components. Future work includes extending \textsf{UniTok} into a general-purpose tokenization interface for foundation models in recommendation.



\section{Acknowledgments}
This work was supported by the National Research Foundation of Korea (NRF) funded by Korea Government (MSIT) under Grants RS-2021-NR059723 and RS-2023-00220762.

\bibliography{aaai2026}

\onecolumn
\appendix

\begin{bibunit}[aaai2026]

\begin{center}
    {\LARGE \bf \textit{Appendix} for ``Tokenize Once, Recommend Anywhere: Unified Item Tokenization for Multi-domain LLM-based Recommendation"}\\[0.5cm]
\end{center}

\vspace{2em}

\section*{List of contents in Appendix}

\begin{itemize}[label=\textcolor{black}{\rule{0.9ex}{0.9ex}}, leftmargin=1.5em]
    \item Summary of \textbf{notations} and their meanings used in the main manuscript
    \item \textbf{Related work} on LLM for generative recommendation and item tokenization: \textbf{Appendix A}
    \item Detailed \textbf{algorithm} of training procedure for \textsf{UniTok}: \textbf{Appendix B}
    \item Formal \textbf{proofs} of the theorems in the main manuscript: \textbf{Appendix C}
    \item \textbf{Additional experimental evaluations}:
    \begin{itemize}[label=\checkmark, leftmargin=2em]
        \item Description and statistics of all \textbf{datasets} used: \textbf{Appendix D.1}
        \item Overview of \textbf{competitors} for comparison: \textbf{Appendix D.2}
        \item Formal definitions of \textbf{evaluation metrics}, including Recall@${\it M}$ and NDCG@${\it M}$: \textbf{Appendix D.3}
        \item \textbf{Implementation details} including hyperparameters and training setup: \textbf{Appendix D.4}
        \item \textbf{Full performance results} across all evaluation metrics: \textbf{Appendix D.5}
        \item \textbf{Ablation study} for evaluation of contributions from individual components: \textbf{Appendix D.6}
        \item \textbf{Sensitivity analysis} on robustness under different hyperparameter settings: \textbf{Appendix D.7}
        \item \textbf{Empirical validation for theorems}: \textbf{Appendix D.8}
    \end{itemize}
\end{itemize}

\vspace{1em}

\begin{center}
    \section*{Table of notations}
\end{center}

\begin{table}[H]
\centering
\small
  \begin{tabular}{cl}
    \toprule
    \bf Notation& \bf Description\\
    \hline
    $\mathcal{D}_k$ & The $k$-th dataset \\
    $K$  & The number of datasets \\
    $\mathcal{I}_k$ & Item set of $\mathcal{D}_k$ \\
    $ \textbf{X}^k $ & Semantic embeddings for domain $k$ \\
    $ d $ & Embedding dimensionality \\
    $\mathbf{x}_i^k$ & Semantic embedding of the $i$-th item in $\mathcal{D}_k$ \\
    $f_\theta $ & Shared encoder \\
    $ g_\phi$ & Shared decoder \\
    $\mathbf{Z}^k $ & Hidden embeddings of $\mathbf{X}^k $ \\
    $\mathbf{z}_i^k$ & Hidden embedding of $\mathbf{x}_i^k$ \\
    $G(\cdot)$ & Router function \\
    $G_k$ & Routing probability assigned to the $k$-th expert \\
    $N $ & Number of selected top experts \\
    $E_k$ & The $k$-th expert module \\
    $E_\text{share}$ & The shared expert module \\
    $L$ & Codebook size \\
    $ T$ & Number of code vector in each codebook \\
    $C_\ell$ & The $\ell$-th codebook \\
    $\mathbf{c}_i^k$ & Discrete token of $\mathbf{x}_i^k$  \\
    $\widehat{\operatorname{HSIC}}(\mathbf{X}^k, \mathbf{Z}^k) $ & Empirical estimate of HSIC between $\textbf{X}^k$ and $\textbf{Z}^k$ \\
    $\mathbf{U}, \mathbf{V}$ & Gaussian kernel matrices \\
    $ \widehat{I}^{(k)}$ & $\widehat{\operatorname{HSIC}}(\mathbf{X}^k, \mathbf{Z}^k) $ \\
    $\operatorname{Var}\left[ \widehat{I}^{(k)} \right]$ & Variance of MI estimates across domains \\
    $\mathbb{E}\left[ \widehat{I}^{(k)} \right]$ & Expectation of MI estimates across domains \\
    $\mathcal{L}_\text{Total}$ & Total loss \\
    $\mathcal{L}_\text{Rec}$ & Reconstruction loss \\
    $\mathcal{L}_\text{RQ} $ & Resdual quantization loss \\
    $\mathcal{L}_\text{MI} $ & Mutual information calibration loss \\
    \bottomrule
    \end{tabular}
\label{tab:notations}
\end{table}

\section{A. Related Work}

Our proposed method is related to two broader areas of research, namely LLM for generative recommendation and item tokenization.

\subsection{1. LLM for Generative Recommendation}


LLMs have been widely adopted in recommender systems for their powerful generative capabilities~\citep{li2024survey, gong2023unified, lin2024data}. They have been utilized not only to enhance traditional recommendation methods but also to directly generate recommendations. A growing body of studies leveraged LLMs for data augmentation by synthesizing user–-item interactions \citep{wei2024llmrec}, incorporating external knowledge \citep{xi2024towards}, and enriching user and item representations at the representation level \citep{liu2024once, qiu2021u, ren2024representation, wu2022userbert}. LLM-based recommender systems employed a projection layer that aligns user and item embeddings within a shared latent space, enabling the computation of similarity scores for ranking tasks \citep{li2023e4srec, lin2024clickprompt, zhang2025collm}. Other methods focused on click-through rate prediction by feeding user profiles and item descriptions into LLMs to model their interactions, predicting the likelihood of user engagement \citep{bao2023tallrec, lin2024rella, prakash2023cr}. However, since LLMs originated from language processing, a gap exists between the language space and the item space when applying them to recommendation tasks.

\subsection{2. Item Tokenization}



In recommender systems powered by LLMs, a fundamental challenge arises from the gap between the language space and the item space. To address this issue, recent research has explored item tokenization strategies that effectively bridge this gap. Existing approaches can be broadly categorized into three types based on the nature of item identifiers. First, ID-based identifiers represent items as discrete tokens using unique IDs \citep{geng2022recommendation, hua2023index, wang2024enhanced}. While simple, these methods lack semantic information and generalization capability. Second, textual identifiers leverage item metadata, such as titles or descriptions, enabling LLMs to utilize pretrained linguistic knowledge for improved semantic understanding and generalization \citep{bao2025bi, dai2023uncovering, li2023generative, liao2023llara, zhang2023recommendation, zhang2021language, lin2024bridging}. Lastly, codebook-based identifiers employ learned discrete representations through vector quantization \citep{rajput2023recommender, wang2024enhanced, zheng2024adapting, wang2024learnable, zheng2024adapting, ZhaiM0YZL025}. These methods combine the interpretability of discrete tokens with the semantic richness of embeddings, providing structured and generalizable representations of items. However, common limitations of these item tokenization techniques hinder the generalization ability of LLMs. Specifically, most approaches rely on training separate models for each dataset \citep{rajput2023recommender, wang2024enhanced, zheng2024adapting, wang2024learnable, zheng2024adapting}, while a more recent approach requires additional costly pre-training to transfer knowledge across domains for adaptation \citep{ZhaiM0YZL025}.

\section{B. Algorithm of \textsf{UniTok}}
\label{app:unitok_alg}
Algorithm~\ref{alg:unitok} outlines the training procedure of the proposed \textsf{UniTok} framework, which unifies item tokenization across multiple domains. Initially (Line 1), all core components, including the encoder, decoder, router, expert modules, and codebooks, are initialized. Each item $i_i^k$ from domain $\mathcal{D}_k$ is embedded into a semantic embedding $\textbf{x}_i^k$ using a pretrained embedding model (Lines 2–6). During training (Lines 7–13), mini-batches are sampled from mixed domains to encourage multi-domain generalization. Each input $\textbf{x}_i^k$ is first encoded into $\textbf{z}_i^k$ (Line 10), then passed through the TokenMoE module to obtain a quantized embedding ${\hat {\bf z}}_i^k$ and discrete token $\textbf{c}_i^k$ via expert routing and residual quantization (Line 11), and finally reconstructed ${\hat {\bf x}_i^k}$ via the decoder (Line 12). The total training objective combines a reconstruction loss (Line 14), a residual quantization loss (Line 15), and a mutual information calibration loss (Line 16) to enforce informativeness and semantic consistency. All parameters are updated jointly (Line 17), and the final output consists of discrete tokens for all items (Line 19).


\begin{algorithm}[tb]
\caption{\textsf{\textbf{UniTok}} Training Procedure}
\label{alg:unitok}
\textbf{Input}: Multi-domain item datasets $\{ \mathcal{D}_1, \mathcal{D}_2, ..., \mathcal{D}_K \}$ with text metadata {\it e.g.}, title, category, description) \\
\textbf{Parameters}: Encoder $f_\theta$, decoder $g_\phi$, router $h(\cdot)$, expert modules $\{ E_1, ..., E_K, E_{\text{share}} \}$ with associated codebooks \\
\textbf{Output}: Discrete tokens $\textbf{c}_i^k \in \mathcal{C}$ for all items
\begin{algorithmic}[1]
\STATE Initialize all modules: $f_\theta$, $g_\phi$, router $h(\cdot)$, experts $E_k$ (including $E_{\text{share}}$), and their codebooks
\FOR{$k = 1$ to $K$}
    \FOR{item $i_i^k \in \mathcal{D}_k$}
        \STATE $\textbf{x}_i^k \leftarrow \text{SemanticEmbedding}(i_i^k)$
    \ENDFOR
\ENDFOR
\FOR{epoch $= 1$ to $E$}
    \STATE Sample mini-batch $\mathcal{B} = \{ \textbf{x}_i^k \}$ from mixed domains
    \FOR{each item $\textbf{x}_i^k$ in batch $\mathcal{B}$}
        \STATE $\textbf{z}_i^k \leftarrow f_\theta(\textbf{x}_i^k)$ \hfill // Encode input
        \STATE $\hat{\textbf{z}}_i^k, \textbf{c}_i^k \leftarrow \text{TokenMoE}(\textbf{z}_i^k)$ \hfill // Discretize via TokenMoE
        \STATE $\hat{\textbf{x}}_i^k \leftarrow g_\phi(\hat{\textbf{z}}_i^k)$ \hfill // Decode reconstruction
    \ENDFOR
    \STATE Compute reconstruction loss $\mathcal{L}_\text{Rec}$ in the main manuscript of Eq. (3)
    \STATE Compute RQ loss $\mathcal{L}_\text{RQ}$ in the main manuscript of Eq. (8)
    \STATE Compute MI loss $\mathcal{L}_\text{MI}$ in the main manuscript of Eq. (10)
    \STATE Update $f_\theta$, $g_\phi$, $h(\cdot)$, and expert parameters
\ENDFOR
\STATE \textbf{return} token assignments $\textbf{c}_i^k$ for all items
\end{algorithmic}
\end{algorithm}



\section{C. Theorems and Proofs}

\textbf{Theorem 1.} \textit{The token space induced by \textsf{UniTok} exhibits strictly higher entropy than that of standard codebook-based methods:}
\[
        H(\mathcal{C}_{\textsf{UniTok}}) > H(\mathcal{C}_{\text{standard}}),
\]
\textit{where $\mathcal{C}_{\text{UniTok}}$ and $\mathcal{C}_{\text{standard}}$ denote the discrete token distributions generated by \textsf{UniTok} and standard codebook-based methods, respectively.}

\textbf{Proof.} Consider the latent space representation of both models:

{\bf Entropy of standard codebook-based methods:} In standard codebook-based methods, a single codebook with $T$ code vectors is used at each of the $L$ levels. The total number of unique tokens in the token space is
\begin{equation}
    |\mathcal{C}_{\text{standard}}| = T^L.
\end{equation}

Assuming a uniform probability distribution over the quantized representations:
\begin{equation}
    P(c) = \frac{1}{T^L},
\end{equation}
The entropy of the token space in standard codebook-based methods is
\begin{equation}
    H(\mathcal{C}_{\text{standard}}) = - \sum_{c \in \mathcal{C}} P(c) \log P(c).
\end{equation}
Substituting $P(c)$, we have
\begin{equation}
    H(\mathcal{C}_{\text{standard}}) = L \log T.
\end{equation}

\textbf{Entropy of \textsf{UniTok}:} In our \textsf{UniTok}, there are $K$ domain-specific experts, each with its own independent codebook of size $T^L$. Given an item embedding $\mathbf{z}$, the router function assigns a probability distribution over experts:
\begin{equation}
    G_k = \frac{\exp(h(\mathbf{z}^{(k)}))}{\sum_{j=1}^{K} \exp(h(\mathbf{z}^{(j)}))},
\end{equation}
where the $h(\cdot)$ is the linear transformation and $\mathbf{z}^{(k)}$ denotes the logit corresponding to the $k$-th domain-specific expert.

Let $\mathcal{G}$ denote the random variable representing expert selection. An item token $c$ in $\mathcal{C}_\text{UniTok}$ is then generated by
\begin{itemize}
    \item Sampling an expert index $k \sim \mathcal{G}$,
    \item Sampling a token $c \sim \mathcal{C}_k$.
\end{itemize}
Here, $\mathcal{C}_k$ is the discrete token distributions in the $k$-th expert.

By the chain rule of entropy, it follows that
\begin{equation}
    H(\mathcal{C}_{\text{UniTok}}) = H(\mathcal{G}) + H(\mathcal{C}_k \mid \mathcal{G}).
\end{equation}

The conditional entropy term $H(\mathcal{C}_k \mid \mathcal{G})$ is the expectation of the entropy of each expert's latent space, weighted by the distribution from the router, and is written as
\begin{equation}
    H(\mathcal{C}_k \mid \mathcal{G}) = \sum_{k=1}^{K} G_k \cdot H(\mathcal{C}_k) = \mathbb{E}_{k \sim \mathcal{G}}[H(\mathcal{C}_k)].
\end{equation}

Then, we have
\begin{equation}
    H(\mathcal{C}_{\text{UniTok}}) = H(\mathcal{G}) + \mathbb{E}_{k \sim \mathcal{G}}[H(\mathcal{C}_k)].
\end{equation}

Since each expert has the same quantization entropy as standard codebook-based methods, it follows that
\begin{equation}
    H(\mathcal{C}_k) = L \log T \quad \text{for all } k,
\end{equation}
We then obtain
\begin{equation}
    \mathbb{E}_{k \sim \mathcal{G}}[H(\mathcal{C}_k)] = L \log T.
\end{equation}

The router entropy $H(\mathcal{G})$ is
\begin{equation}
    H(\mathcal{G}) = - \sum_{k=1}^{K} G_k \log G_k,
\end{equation}
which is always positive when $K>1$, i.e., $H(\mathcal{G}) > 0$.

Thus, the entropy of $\mathcal{C}_\textsf{UniTok}$ is
\begin{equation}
    H(\mathcal{C}_{\textsf{UniTok}}) = H(\mathcal{G}) + L \log T > H(\mathcal{C}_{\text{standard}}).
    \label{eq:entropy_gain}
\end{equation}

This completes the proof of Theorem 1.\hfill $\blacksquare$

\vspace{1em}

Before proving Theorem 2, we begin with Lemma 1 below.

\vspace{1em}

\textbf{Lemma 1.} \textit{The function $f(\mathbf{x}) = \|\mathbf{x}\|^2 = \mathbf{x}^\top \mathbf{x}$, where $\mathbf{x} \in \mathbb{R}^n$, is convex.}

\textbf{Proof.} To show that $f(\mathbf{x})$ is convex, we check whether its Hessian matrix is positive semidefinite.

\textbf{Step 1: Compute the gradient of $f(\mathbf{x})$}  as follows:
\begin{equation}
    \nabla f(\mathbf{x}) = \nabla (\mathbf{x}^\top \mathbf{x}) = 2\mathbf{x}.
\end{equation}

\textbf{Step 2: Compute the Hessian of $f(\mathbf{x})$}  as follows:
\begin{equation}
    \nabla^2 f(\mathbf{x}) = \nabla (2\mathbf{x}) = 2\mathbf{I},
\end{equation}
where $\mathbf{I}$ is the $n \times n$ identity matrix.

Therefore, the Hessian is positive definite, which implies that $f(\mathbf{x})$ is a strictly convex function. This completes the proof of Lemma 1.
\hfill $\blacksquare$

\vspace{1em}

\textbf{Theorem 2.} \textit{Let $\mathbb{E}[\mathcal{L}_\textsf{UniTok}]$ and $\mathbb{E}[\mathcal{L}_{\text{standard}}]$ denote the expected quantization error of \textsf{UniTok} and standard codebook-based methods using a single shared codebook, respectively. Then, the following inequality holds:}
\[
    \mathbb{E}[\mathcal{L}_{\textsf{UniTok}}] \leq \mathbb{E}[\mathcal{L}_{\text{standard}}].
\]

\textbf{Proof.} Let $\mathbf{z} \in \mathbb{R}^d$ be an item latent embedding drawn from a distribution $p(z)$. Let $\mathbf{{\hat z}} = E(\mathbf{z})$ denote the output of the quantization function, which generates a combination of code vectors to approximate the item latent embedding $\mathbf{z}$.


For standard codebook-based methods using a single set of codebooks, the expected quantization error is given by

\begin{equation}
\mathbb{E}[\mathcal{L}_{\text{standard}}] = \mathbb{E}_{\mathbf{z} \sim p(z)} \left[ \| \mathbf{z} - E(\mathbf{z}) \|^2 \right].
\end{equation}

Now, consider \textsf{UniTok} with $K$ domain-specific experts, each with its own set of codebooks. Given the embedding $\mathbf{z}$, the router function that assigns probabilities to each expert:
\begin{equation}
G_k = \frac{\exp(h(\mathbf{z}^{(k)}))}{\sum_{j=1}^{K} \exp(h(\mathbf{z}^{(j)}))}.
\end{equation}



The approximated latent embedding is generated by $\mathbf{{\hat z}}=\sum_{k=1}^{K} G_k \cdot E_k(\mathbf{z})$. Thus, the quantization error is
\begin{equation}
\mathcal{L}_{\textsf{UniTok}}(\mathbf{z}) =  \left\| \mathbf{z} - \sum_{k=1}^{K} G_k \cdot E_k(\mathbf{z}) \right\|^2.
\end{equation}
where $E_k(\mathbf{z})$ is the quantization function of the $k$-th expert module. As $\sum_{k=1}^{K} G_k = 1$, we have
\begin{equation}
\mathcal{L}_{\textsf{UniTok}}(\mathbf{z}) =  \left\| \sum_{k=1}^{K} G_k \cdot (\mathbf{z} -  E_k(\mathbf{z})) \right\|^2.
\end{equation}





We define the approximation error for each $z$ as
\begin{equation}
\epsilon_k(\mathbf{z}) = \mathbf{z} - E_k(\mathbf{z}).
\end{equation}

Replacing with $\epsilon_k(\mathbf{z})$, we have
\begin{equation}
\mathcal{L}_{\textsf{UniTok}}(\mathbf{z}) = \left\|\sum_{k=1}^{K} G_k \cdot \epsilon_k(\mathbf{z}) \right\|^2.
\end{equation}

Using Jensen's inequality, we apply the convexity of the squared norm function based on Lemma 1:
\begin{equation}
\mathcal{L}_{\textsf{UniTok}}(\mathbf{z}) = \left\| \sum_{k=1}^{K} G_k \cdot \epsilon_k(\mathbf{z}) \right\|^2 \leq \sum_{k=1}^{K} G_k \cdot \| \epsilon_k(\mathbf{z}) \|^2.
\end{equation}

Taking the expectation over the data distribution $p(z)$ on both sides of Eq. (21), we obtain
\begin{equation}
\mathbb{E}_{\mathbf{z} \sim p(z)} \left[ \left\| \sum_{k=1}^{K} G_k \cdot \epsilon_k(\mathbf{z}) \right\|^2 \right] \leq \mathbb{E}_{\mathbf{z} \sim p(z)} \left[ \sum_{k=1}^{K}G_k \cdot \| \epsilon_k(\mathbf{z}) \|^2 \right].
\end{equation}

Since the single quantization function $E(\mathbf{z})$ can be viewed as a degenerate case where $G_k = 1$ for a single expert and $0$ for all others, the \textsf{UniTok} always provides a better or equal approximation:
\begin{equation}
\mathbb{E}_{\mathbf{z} \sim p(z)} \left[ \mathcal{L}_{\textsf{UniTok}}(\mathbf{z}) \right] \leq \mathbb{E}_{\mathbf{z} \sim p(z)}[\mathcal{L}_{\text{standard}}(\mathbf{z})].
\end{equation}



This completes the proof of Theorem 2.
\hfill $\blacksquare$

\vspace{1em}

\textbf{Theorem 3.}
\textit{Suppose that the loss $\mathcal{L}^{(k)}$ on the $k$-th domain is Lipschitz-continuous with respect to the informativeness of representations. Then, the performance variability across domains is upper-bounded by the variance of MI:}
\[
    |\mathcal{L}^{(i)} - \mathcal{L}^{(j)}| \leq C \sqrt { \text{Var}\left[ {\widehat{I}^{(k)}} \right]}, \forall i,j
\]
\textit{where $\text{Var}\left[ {\widehat{I}^{(k)}} \right]$ is the variance of MI estimates across domains and $C$ is a constant.}

\textbf{Proof.} Let $\widehat{I}^{(k)} = \widehat{\mathrm{HSIC}}(\mathbf{X}^k, \mathbf{Z}^k)$ be the MI between the input $\mathbf{X}^k$ and its representation $\mathbf{Z}^k$ for the $k$-th domain. The loss $\mathcal{L}^{(k)}$ on the $k$-th domain is Lipschitz-continuous with respect to the informativeness of representation, {\it i.e.},
\begin{equation}
    \left| {\mathcal{L}^{\left( i \right)}  - \mathcal{L}^{\left( j \right)} } \right| \le L \left| \widehat{\mathrm{HSIC}}{\left( {\mathbf{X}^i ;\mathbf{Z}^i } \right) - \widehat{\mathrm{HSIC}}\left( {\mathbf{X}^j ;\mathbf{Z}^j } \right)} \right|,
\end{equation}
where $L$ is the Lipschitz constant that quantifies the sensitivity of the downstream loss to changes in representation informativeness (i.e., MI).

This assumption ensures that variations in representation informativeness (measured by MI) translate into bounded variations in the downstream task loss, effectively connecting the input--output MI stability \citep{xu2017information}. As a result, controlling the variance of MI across domains allows us to stabilize performance fluctuations.

Now, we take the maximum gap in MI across all domain pairs and relate it to the variance of MI:
\begin{equation}
    \begin{array}{l}
     \;\;\;\left| {\widehat{I}^{(i)} - \widehat{I}^{(j)}} \right| \\ 
     \mathop \le \max \limits_{i,j} \left| {\widehat{I}^{(i)} - \widehat{I}^{(j)}} \right| \\ 
     = \max \limits_{i,j} \left| {\widehat{I}^{(i)} - \mu - (\widehat{I}^{(j)} - \mu)} \right| \\
     \le \max \limits_{i,j} (\left| {\widehat{I}^{(i)} - \mu }\right|+ \left| {\left(\widehat{I}^{(j)} - \mu\right)} \right|)\\
      \le 2 \mathop {\max }\limits_k \left| {\widehat{I}^{(k)} - \mu } \right| \\ 
      \le 2 \sqrt{\sum\limits_{k=1}^{K}\left(\widehat{I}^{(k)}-\mu\right)^2} \\
      \le 2\sqrt {K Var\left[ {\widehat{I}^{(k)}} \right]}.  \\ 
    \end{array}
\end{equation}

Therefore, we obtain
\[
\left| {\mathcal{L}^{\left( i \right)}  - \mathcal{L}^{\left( j \right)} } \right| \le L \left| \widehat{I}^{(i)} - \widehat{I}^{(j)} \right| \le C\sqrt {Var\left[ {\widehat{I}^{(k)}} \right]},
\]
where $C=2L\sqrt{K}$.

This completes the proof of Theorem 3.
\hfill $\blacksquare$

\section{D. Additional Experimental Evaluations}

\subsection{1. Datasets}

 We conduct our experiments on ten real-world datasets widely adopted for evaluating the performance of recommendations, which include Beauty, Cellphones, Grocery, Instruments, Office, Pet Supplies, Tools, Toys, Games\footnote{https://nijianmo.github.io/amazon/index.html.}, and Yelp\footnote{https://www.yelp.com/dataset.}. Each item in these datasets contains rich content information, including title, category, and description. Additionally, we use three datasets,  Clothing, Health, and Sports\footnote{https://nijianmo.github.io/amazon/index.html.}, for zero-shot evaluation. Table \ref{table:datasets} provides a summary of the statistics for all datasets.

\begin{table}[h]
\small
\centering
\setlength\tabcolsep{4pt}
  \begin{tabular}{cccccl}
    \toprule
    {\bf Dataset} & {\bf \# of users} & {\bf \# of items} & {\bf \# of interactions} \\
    \midrule
    {\bf Beauty} & 22,363 & 12,101 & 198,502 \\
    {\bf Cellphones} & 27,879 & 10,429 & 194,439 \\
    {\bf Grocery} & 14,681 & 8,713 & 151,254  \\
    {\bf Instruments} & 24,772 & 9,922 & 206,153 \\
    {\bf Office Products} & 4,905 & 2,420 & 53,258 \\
    {\bf Pet Supplies} & 19,856 & 8,510 & 157,836 \\
    {\bf Tools} & 16,638 & 10,217 & 134,476 \\
    {\bf Toys} & 19,412 & 11,924 & 167,597 \\
    {\bf Games} & 24,303 & 10,672 & 231,780 \\
    {\bf Yelp} & 30,431 & 20,033 & 316,354 \\
    {\bf Clothing} & 39,387 & 23,033 & 278,677 \\
    {\bf Health} & 38609 & 18533 & 346,355 \\
    {\bf Sports} & 35,598 & 18,357 & 296,337 \\
  \bottomrule
\end{tabular}
\caption{The statistics of datasets from ten domains suited to recommendations.}
\label{table:datasets}
\end{table}

\subsection{2. Competitors}

To comprehensively demonstrate the superiority of \textsf{UniTok}, we present nine benchmark recommendation methods, including four widely-used collaborative filtering methods, MF \citep{rendle2012bpr}, SASRec \citep{kang2018self}, LightGCN \citep{he2020lightgcn}, Bert4Rec \citep{sun2019bert4rec}, and five item tokenization-aided recommendations, P5-TID \citep{hua2023index}, P5-SemID~\citep{hua2023index}, TIGER~\citep{rajput2023recommender}, LC-Rec~\citep{zheng2024adapting}, and LETTER~\citep{wang2024learnable}. We implemented all these methods using the parameter settings described in their original articles.
\begin{itemize}
    \item \textbf{MF~\citep{rendle2012bpr}}. This models user-item interactions by decomposing them into latent user and item embeddings through matrix factorization.

    \item \textbf{SASRec~\citep{kang2018self}}. This method uses self-attention to model user sequences, effectively capturing both short- and long-term dependencies with high efficiency.

    \item \textbf{LightGCN~\citep{he2020lightgcn}}. This model simplifies graph convolutional networks (GCNs) for recommendation by focusing solely on neighborhood aggregation, eliminating feature transformation and nonlinear activation to enhance performance.

    \item \textbf{Bert4Rec~\citep{sun2019bert4rec}}. This method leverages bidirectional self-attention with a Cloze-style objective to model user behavior sequences, enabling richer context modeling than unidirectional sequential methods.

    \item \textbf{P5-TID~\citep{hua2023index}}. This method uses item titles as textual identifiers to enable LLM-based generative recommendation.

    \item \textbf{P5-SemID~\citep{hua2023index}}. This method constructs item identifiers from metadata (e.g., attributes) for LLM-based generative recommendation.

    \item \textbf{TIGER~\citep{rajput2023recommender}}. This method trains a sequence-to-sequence model to autoregressively generate item identifiers composed of semantic codeword tuples for next-item prediction.

    \item \textbf{LC-Rec~\citep{zheng2024adapting}}. This method integrates language and collaborative semantics by learning meaningful item indices via vector quantization and tuning LLMs through alignment tasks for direct item generation.

    \item \textbf{LETTER~\citep{wang2024learnable}}. This method is a learnable tokenizer for LLM-based generative recommendation that combines residual quantization, contrastive alignment, and diversity loss to encode hierarchical semantics, capture collaborative signals, and mitigate code assignment bias.
\end{itemize}

\subsection{3. Performance Metrics}

We follow the \textbf{full-ranking evaluation protocol}~\citep{he2020lightgcn}, where all non-interacted items are ranked for each user. To evaluate performance, we adopt two widely used ranking metrics: \textbf{Recall@{\em M} (R@{\em M})} and \textbf{NDCG@{\em M} (N@{\em M})}, where $M \in \{5, 10\}$. We formally define the ranking metrics used in our experiments.

{\bf Recall@{\em M}.} Recall@{\it M} measures the proportion of relevant items successfully retrieved in the top-$M$ recommendations:
\[
\text{Recall@{\it M}} = \frac{|\text{Top-}M \ \text{recommended items} \cap \text{Ground truth ttems}|}{|\text{Ground truth items}|}.
\]

{\bf NDCG@{\em M} (Normalized Discounted Cumulative Gain).} NDCG@{\em M} considers both the relevance and the ranking positions of the recommended items:
\[
\text{NDCG@{\it M}} = \frac{1}{\text{IDCG@{\it M}}} \sum_{i=1}^{M} \frac{\mathds{1}\{\text{item}_i \in \text{Ground truth}\}}{\log_2(i+1)},
\]
where $\text{IDCG@{\it M}}$ denotes the ideal DCG, i.e., the maximum possible DCG for the given ground truth, ensuring normalization between 0 and 1.

\subsection{4. Implementation Details}

To evaluate recommendation performance, we adopt TIGER \citep{rajput2023recommender}, a representative LLM-based generative recommender system that uses T5 \citep{raffel2020exploring} as its backbone. We follow this design and retain T5 in our experiments due to its unified text-to-text framework, strong performance across a wide range of generative tasks, and efficiency in conditional sequence generation—making it well-suited to the generative recommendation paradigm adopted by TIGER.

\begin{table*}[h]\centering
\setlength\tabcolsep{5pt}
\small
  \begin{tabular}{c|cc|cc|cc|cc|cc}
    \toprule[1pt]
    \multicolumn{1}{c|}{}&\multicolumn{2}{|c|}{{\bf Beauty}}&\multicolumn{2}{c|}{{\bf Cellphones}}&\multicolumn{2}{c}{{\bf Grocery}}&\multicolumn{2}{|c|}{{\bf Insturments}}&\multicolumn{2}{|c}{{\bf Office Products}}\\
    \cmidrule{1-11}
           {\bf Method} & {\bf R@10} & {\bf N@10}& {\bf R@10}& {\bf N@10}& {\bf R@10}& {\bf N@10}& {\bf R@10}& {\bf N@10}& {\bf R@10}& {\bf N@10}\\
    \midrule[1pt]
    {\bf MF} & 0.0614& 0.0369 & 0.0604& 0.0267 & 0.0418& 0.0216 & 0.0930& 0.0710 & 0.0569&0.0255 \\
    {\bf LightGCN} & 0.0639& 0.0285 & 0.0668& 0.0456 & 0.0697 & 0.0357 &0.1008 & 0.0781 & 0.0587& 0.0301 \\
    {\bf SASRec} & 0.0646 & 0.0314 &0.0651 & 0.0446 & 0.0701& 0.0376 &0.0905 & 0.0609 & 0.0574 & 0.0285 \\
    {\bf Bert4Rec} & 0.0372& 0.0194 & 0.0507& 0.0268 & 0.0448& 0.0237 & 0.0791& 0.0573 &0.0563 & 0.0274 \\
    \midrule
    {\bf P5-TID} & 0.0532 & 0.0255 & 0.0648& 0.0357 & 0.0617& 0.0316 & 0.0928& 0.0721 &0.0557 & 0.0239 \\
    {\bf P5-SemID} & 0.0584& 0.0304 & 0.0737& 0.0406 & 0.0641& 0.0351 & 0.0964& 0.0730 &0.0592 & 0.0283\\
    {\bf TIGER} & 0.0624 & 0.0324 & 0.0838 & 0.0446 & 0.0706 & 0.0375 & 0.1047 & 0.0788 & 0.0594 & 0.0295 \\
    {\bf LC-Rec} & \underline{0.0684} & \underline{0.0381}& 0.0859& 0.0458 &0.0722 & 0.0369 &0.1066 & 0.0802 & 0.0637& 0.0311 \\
    {\bf LETTER}& 0.0672 & 0.0364& \underline{0.0876}& \underline{0.0473}& \underline{0.0731}& \underline{0.0392} & \underline{0.1122}& \underline{0.0831} &\underline{0.0649} & \underline{0.0326} \\
    \rowcolor{lightgray!30}
    {\bf \textsf{UniTok}}& {\bf 0.0934}& {\bf 0.0478}& {\bf 0.1251}& {\bf 0.0647}& {\bf 0.1061}& {\bf 0.0533}& {\bf 0.1361}& {\bf 0.0884} &{\bf 0.0897} & {\bf 0.0432}\\
    \midrule
    {\bf Improve}& 36.55\% & 25.46\% & 42.81\% & 36.78\% & 45.14\% & 35.97\% & 21.30\% & 6.38\% & 38.21\% & 32.52\% \\
    \midrule[1pt]
        \multicolumn{1}{c|}{}&\multicolumn{2}{|c|}{{\bf Pet Supplies}}&\multicolumn{2}{|c|}{{\bf Tools}}&\multicolumn{2}{|c|}{{\bf Toys}}&\multicolumn{2}{|c|}{{\bf Games}}&\multicolumn{2}{|c}{{\bf Yelp}}\\
    \cmidrule{1-11}
           {\bf Method} & {\bf R@10}& {\bf N@10}& {\bf R@10}& {\bf N@10}& {\bf R@10}& {\bf N@10}& {\bf R@10}& {\bf N@10}& {\bf R@10}& {\bf N@10}\\
    \midrule[1pt]
    {\bf MF} & 0.0503& 0.0268 & 0.0356& 0.0169 & 0.0405& 0.0192 & 0.0359& 0.0366 & 0.0304& 0.0144\\
    {\bf LightGCN} & 0.0529& 0.0289 & 0.0482& 0.0257 & 0.0495& 0.0287 & 0.0407& 0.0417 & 0.0368 & 0.0195\\
    {\bf SASRec} & 0.0538 & 0.0301 & 0.0475& 0.0234 & 0.0473& 0.0239 & 0.0401 & 0.0412 & 0.0354& 0.0183\\
    {\bf Bert4Rec} & 0.0313& 0.0161 &0.0184 & 0.0092 & 0.0333&0.0177 & 0.0363& 0.0379 &0.0272 & 0.0131\\
    \midrule
    {\bf P5-TID} & 0.0485& 0.0243 & 0.0377& 0.0198 & 0.0419& 0.0202 &0.0372 & 0.0388 & 0.0316& 0.0154\\
    {\bf P5-SemID} & 0.0569 & 0.0282 &0.0405 & 0.0237 &0.0445 & 0.0231 & 0.0398& 0.0432 & 0.0324& 0.0188\\
    {\bf TIGER} & 0.0546 & 0.0279 & 0.0507 & 0.0284 & 0.0486 & 0.0268 & 0.0438 & 0.0427 & 0.0394 & 0.0208 \\
    {\bf LC-Rec} & \underline{0.0648}& \underline{0.0335} & \underline{0.0561}& \underline{0.0307} & 0.0538& 0.0279 &0.0442 & 0.0451 & 0.0418& 0.0215\\
    {\bf LETTER}& 0.0596 & 0.0307 & 0.0556 & 0.0298 & \underline{0.0546} & \underline{0.0291} & \underline{0.0559} & \underline{0.0469} & \underline{0.0426} & \underline{0.0231}\\
    \rowcolor{lightgray!30}
    {\bf \textsf{UniTok}} & {\bf 0.0955} & {\bf 0.0496} & {\bf 0.0852} & {\bf 0.0439} & {\bf 0.0902} & {\bf 0.0442} & {\bf 0.0565} & {\bf 0.0476} & {\bf 0.0684} & {\bf 0.0321}\\
    \midrule
    {\bf Improve}& 47.38\% & 48.06\% & 51.87\% & 42.99\% & 65.20\% & 51.89\% & 1.07\% & 1.49\% & 60.56\% & 38.96\% \\
    \bottomrule[1pt]
  \end{tabular}
    \caption{Performance comparison among \textsf{UniTok} and recommendation competitors for the ten benchmark datasets. Here, the best and second-best performers are highlighted by bold and underline, respectively. The improvements are statistically significant on average ($p = 0.0219 < 0.05$) based on paired t-tests over five runs across all datasets.}
    \label{tab:app_overall}
\end{table*}

While other LLMs ({\it e.g.}, GPT, BART, and LLaMA) are viable options, our focus is to ensure a fair and controlled evaluation of our proposed tokenization framework, \textsf{UniTok}. Using T5, consistent with other codebook based methods like TIGER \citep{rajput2023recommender} and LETTER \citep{wang2024learnable}, allows us to attribute performance improvements to our method rather than differences in model architecture. Additionally, T5 offers a favorable balance between capability and computational efficiency, which is important for reproducible experimentation. We leave the exploration of other LLMs as promising future work.

For item tokenization, we use 4-level codebook-based identifiers, where each codebook comprises 256 code vectors with an embedding dimension of 32. To obtain item semantic embeddings, we follow \citep{rajput2023recommender, wang2024learnable} and use a pre-trained language model to encode item content information. The resulting semantic embedding has a dimensionality of 2048, while the hidden embedding dimension is 32. The shared expert is initialized with the average item embedding across all domains, capturing general semantic patterns over domains, whereas the $K=10$ domain experts---corresponding to ten domains---are initialized with their respective domain-specific average embeddings, naturally inducing domain-aware assignment without supervision. Following \citep{rajput2023recommender, wang2024learnable}, we set $\alpha$ as 0.25 and $\beta$ as 1. And $\lambda_\text{RQ}$ is typically set to 1 and $\lambda_\text{MI}$ in the range of $\left\{0.01, 0.03, 0.05, 0.07, 0.1\right\}$. We train \textsf{UniTok} for 10k epochs via AdamW [27] optimizer with a learning rate of $1e-3$ and a batch size of $1, 024$. TIGER is fine-tuned for convergence based on the validation performance, with a learning rate in {1e-3, 5e-4} and {1e-4, 2e-4, 3e-4}. 

All experiments are carried out with Intel (R) 12-Core (TM) E5-1650 v4 CPUs @ 3.60GHz and NVIDIA GeForce RTX 3090 GPUs.

\subsection{5. Complete Set of Experimental Results}

To evaluate the recommendation accuracy of \textsf{UniTok}, we compare it with baseline tokenization methods across multiple benchmark datasets, measuring performance using both Recall@10 and NDCG@10 (note that we have shown the results only with respect to NDCG@10 in the main manuscript due to space limit). Unlike existing approaches that train a separate tokenizer per dataset, \textsf{UniTok} is trained once across all domains. As shown in Table~\ref{tab:app_overall}, \textsf{UniTok} consistently outperforms competitors, achieving up to 65.20\% improvement in Recall@10 on the Toys dataset, as confirmed by paired t-tests showing statistically significant improvements ($p < 0.05$). The results demonstrate the effectiveness of our unified tokenization framework; its shared tokenization captures item semantics across diverse domains without domain-specific customization, highlighting the model’s scalability and generality.

LLM-based recommender systems using item tokenization, such as TIGER, LC-Rec, LETTER, and \textsf{UniTok}, consistently outperform traditional collaborative filtering methods ({\it e.g.}, MF, LightGCN, SASRec, and Bert4Rec), benefiting from LLMs’ semantic reasoning capabilities. Moreover, incorporating learned tokenization yields further gains over metadata-based approaches ({\it e.g.}, P5-TID and P5-SemID), as tokenization bridges the item and language domains, enabling discrete, semantically rich item representations that enhance generalization.

\subsection{6. Ablation Study}

To assess the contribution of each component in \textsf{UniTok}, we perform an ablation study by progressively removing or modifying its core modules: the TokenMoE module, the shared expert, and the MI calibration part. \textsf{UniTok-1} removes the TokenMoE and MI calibration; \textsf{UniTok-2} keeps TokenMoE, but removes the shared expert and MI calibration; and \textsf{UniTok-3} only removes the MI calibration (note that we have shown the results only on three datasets in the main manuscript due to space limit). \textsf{UniTok} includes all components. As shown in Table~\ref{tab:ablation_main}, removing any module leads to a noticeable drop in recommendation accuracy, which confirms that the combination of modules is crucial to \textsf{UniTok}’s effectiveness. In particular, comparing \textsf{UniTok-1} and \textsf{UniTok-2} highlights the importance of the TokenMoE module, which significantly improves performance across all datasets by capturing dataset-specific token semantics. Additionally, in comparison with \textsf{UniTok-3}, introducing MI calibration further enhances performance by encouraging the learned representations to retain essential semantic information from the original input space. For clarity and focus, we report results on five representative domains that exhibit trends consistent with those observed across the full range of evaluated datasets.


\begin{table}[h]\centering
\setlength\tabcolsep{7pt}
\small
  \begin{tabular}{c|cc|cc|cc|cc|cc}
    \toprule[1pt]
    \multicolumn{1}{c|}{}&\multicolumn{2}{|c|}{{\bf Beauty}}&\multicolumn{2}{c|}{{\bf Cellphones}}&\multicolumn{2}{c|}{{\bf Grocery}} &\multicolumn{2}{c|}{{\bf Instruments}} &\multicolumn{2}{c}{{\bf Yelp}}\\
    \cmidrule{1-11}
           {\bf Method} & {\bf R@10}& {\bf N@10}& {\bf R@10}& {\bf N@10}& {\bf R@10}& {\bf N@10} & {\bf N@10}& {\bf R@10} & {\bf N@10}& {\bf R@10}\\
    \midrule[1pt]
    {\bf \textsf{UniTok-1}}& 0.0558 & 0.0304 &  0.0702& 0.0371 & 0.0633 & 0.0342 & 0.0926 & 0.0742 & 0.0345 & 0.0177\\
    {\bf \textsf{UniTok-2}}& 0.0896 & 0.0436 & 0.1194 & 0.0606 & 0.0989 & 0.0497 & 0.1273 & 0.0851 & 0.0624 & 0.0281\\
    {\bf \textsf{UniTok-3}}& 0.0915& 0.0457& 0.1225& 0.0622& 0.1044& 0.0515 & 0.1327& 0.0868& 0.0657& 0.0303\\
    \rowcolor{lightgray!30}
    {\bf \textsf{UniTok}}& {\bf 0.0934}& {\bf 0.0478}& {\bf 0.1251}& {\bf 0.0647}& {\bf 0.1061}& {\bf 0.0533} & {\bf 0.1361}& {\bf 0.0884}& {\bf 0.0684}& {\bf 0.0321}\\
    \bottomrule[1pt]
  \end{tabular}
    \caption{Ablation study results on the Beauty, Cellphones, Grocery, Instrument, and Yelp datasets.}
  \label{tab:ablation_main}
\end{table}

\subsection{7. How Sensitive is \textsf{UniTok} to Key Parameters?}

\begin{figure}[h]
    \centering
    \begin{subfigure}[b]{0.22\linewidth} 
        \centering
        \includegraphics[width=\linewidth]{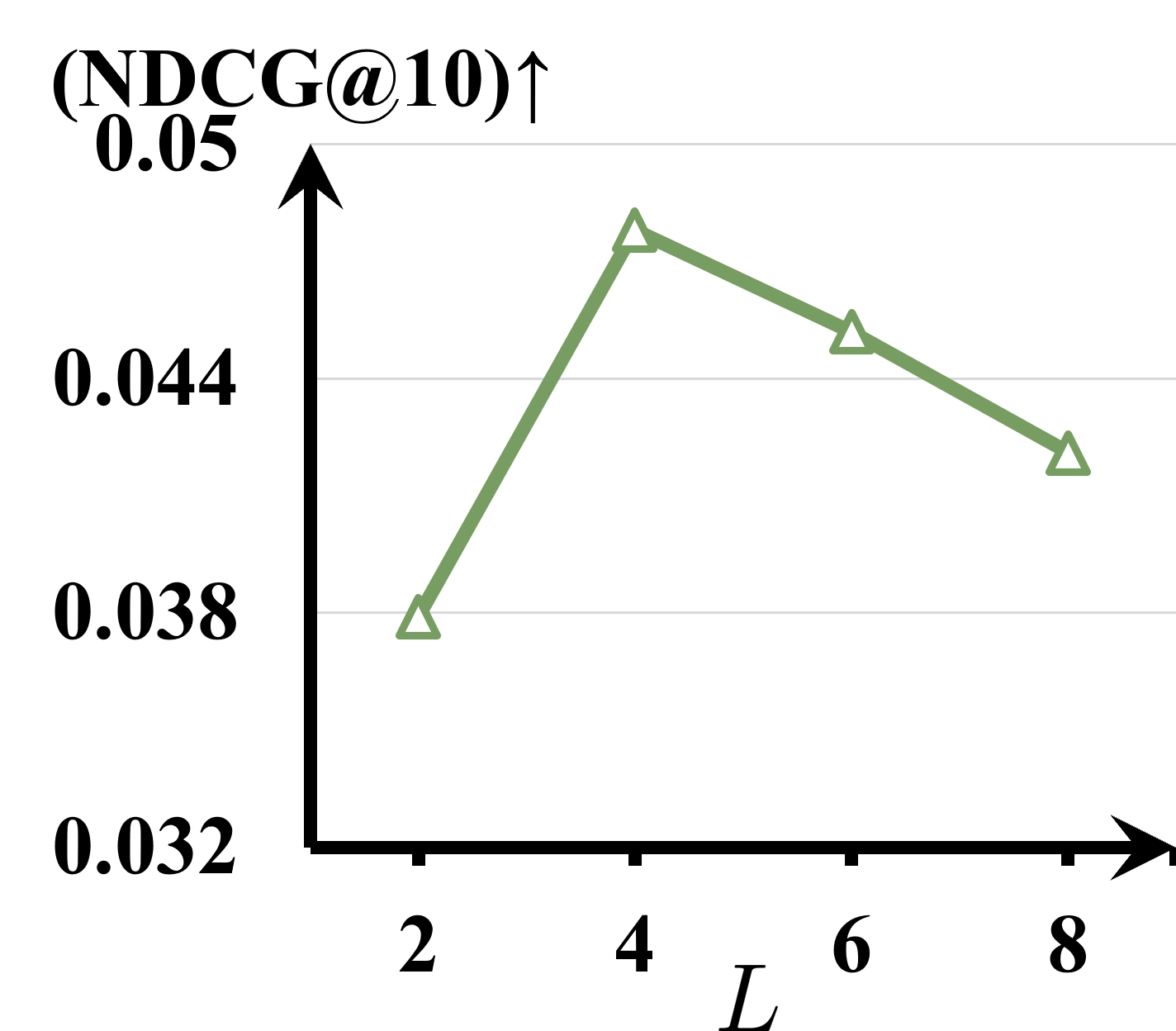} 
        \captionsetup{font={small,stretch=0.5}, skip=5pt,textfont=normalfont}
        \caption{}
        \label{fig:L_beauty}
    \end{subfigure}
     \begin{subfigure}[b]{0.22\linewidth} 
        \centering
        \includegraphics[width=\linewidth]{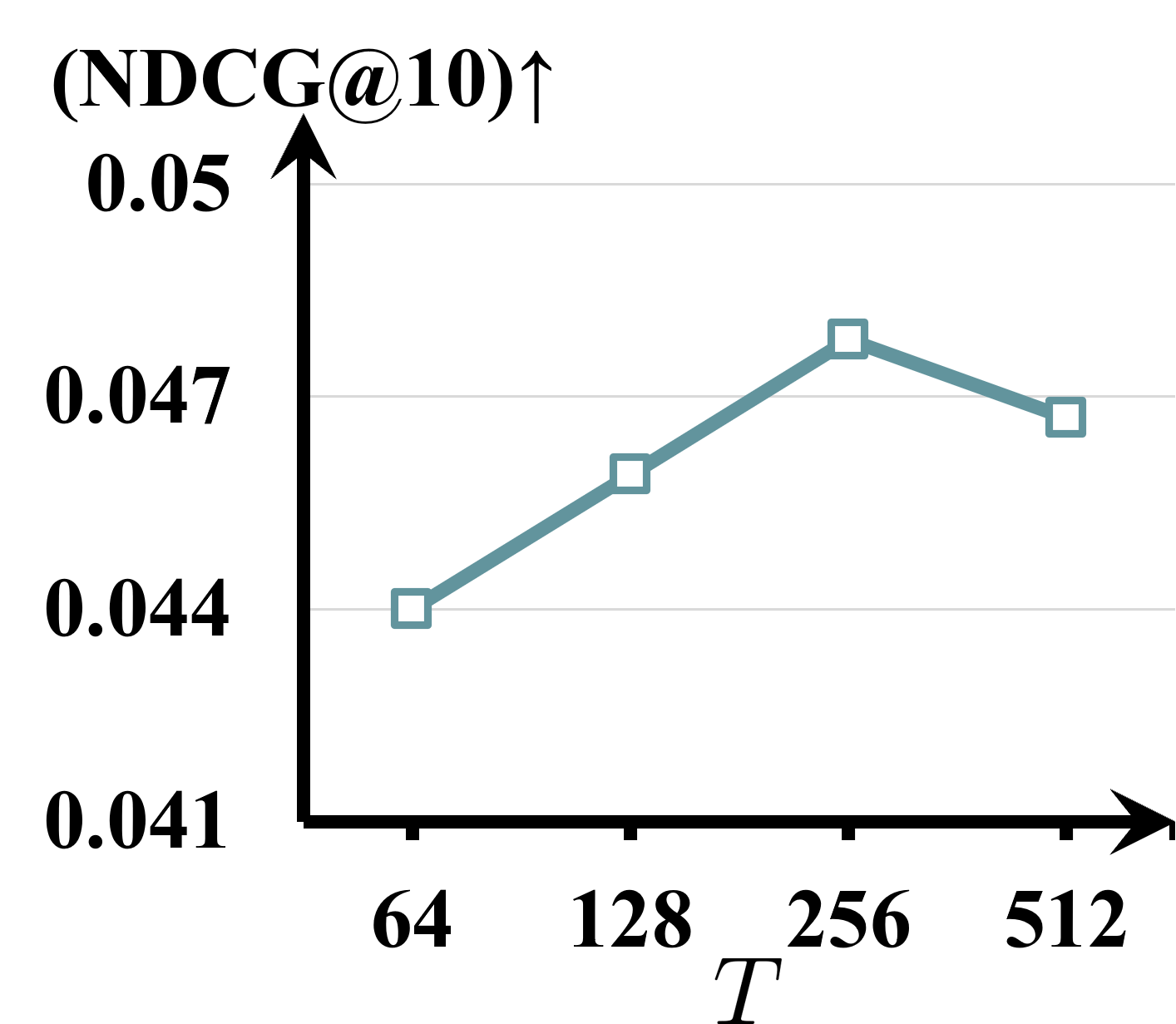} 
        \captionsetup{font={small,stretch=0.5}, skip=5pt,textfont=normalfont}
        \caption{}
        \label{fig:T_beauty}
    \end{subfigure}
    \begin{subfigure}[b]{0.22\linewidth} 
        \centering
        \includegraphics[width=\linewidth]{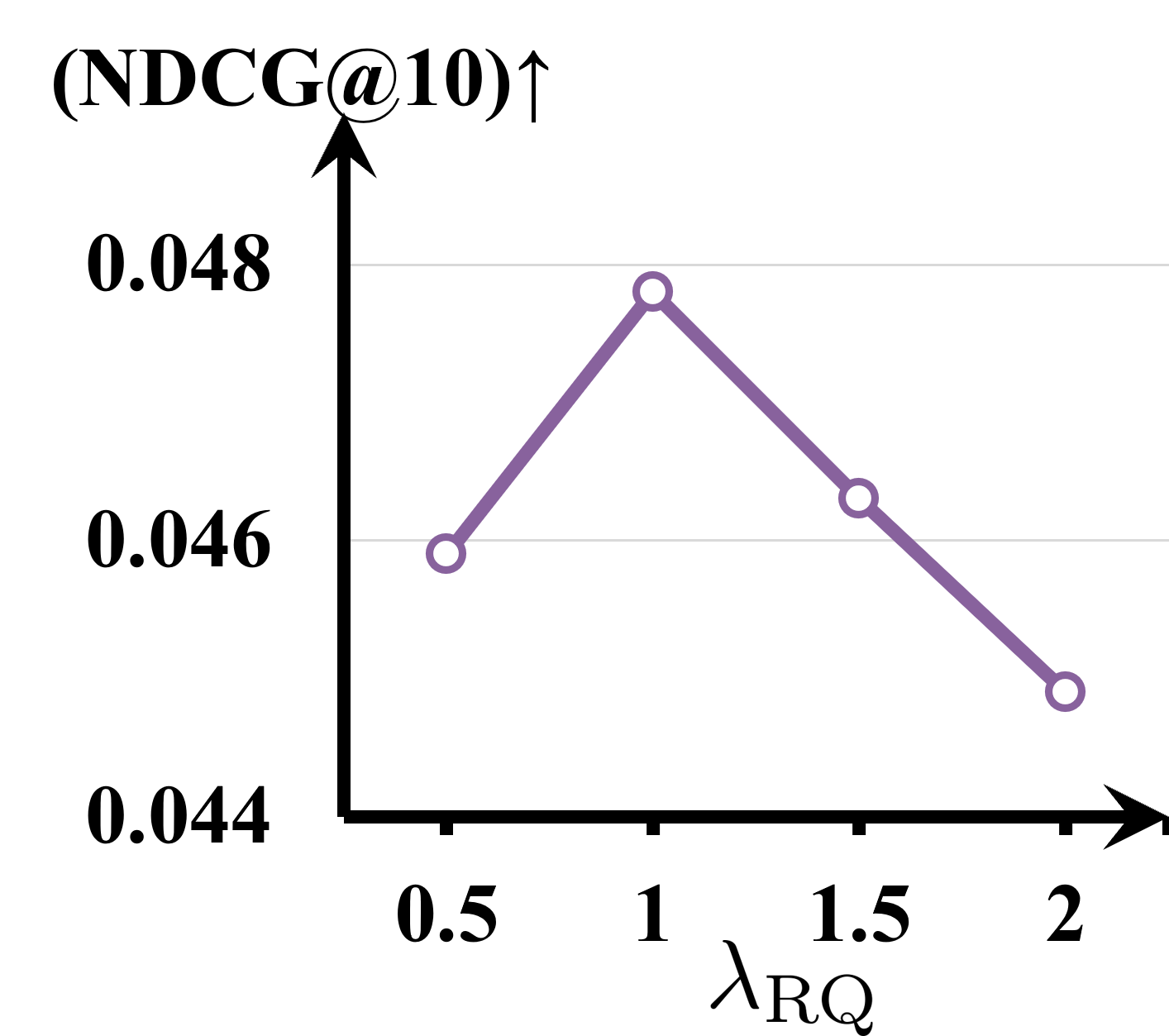} 
        \captionsetup{font={small,stretch=0.5}, skip=5pt,textfont=normalfont}
        \caption{}
        \label{fig:RQ_beauty}
    \end{subfigure}
    \begin{subfigure}[b]{0.22\linewidth} 
        \centering
        \includegraphics[width=\linewidth]{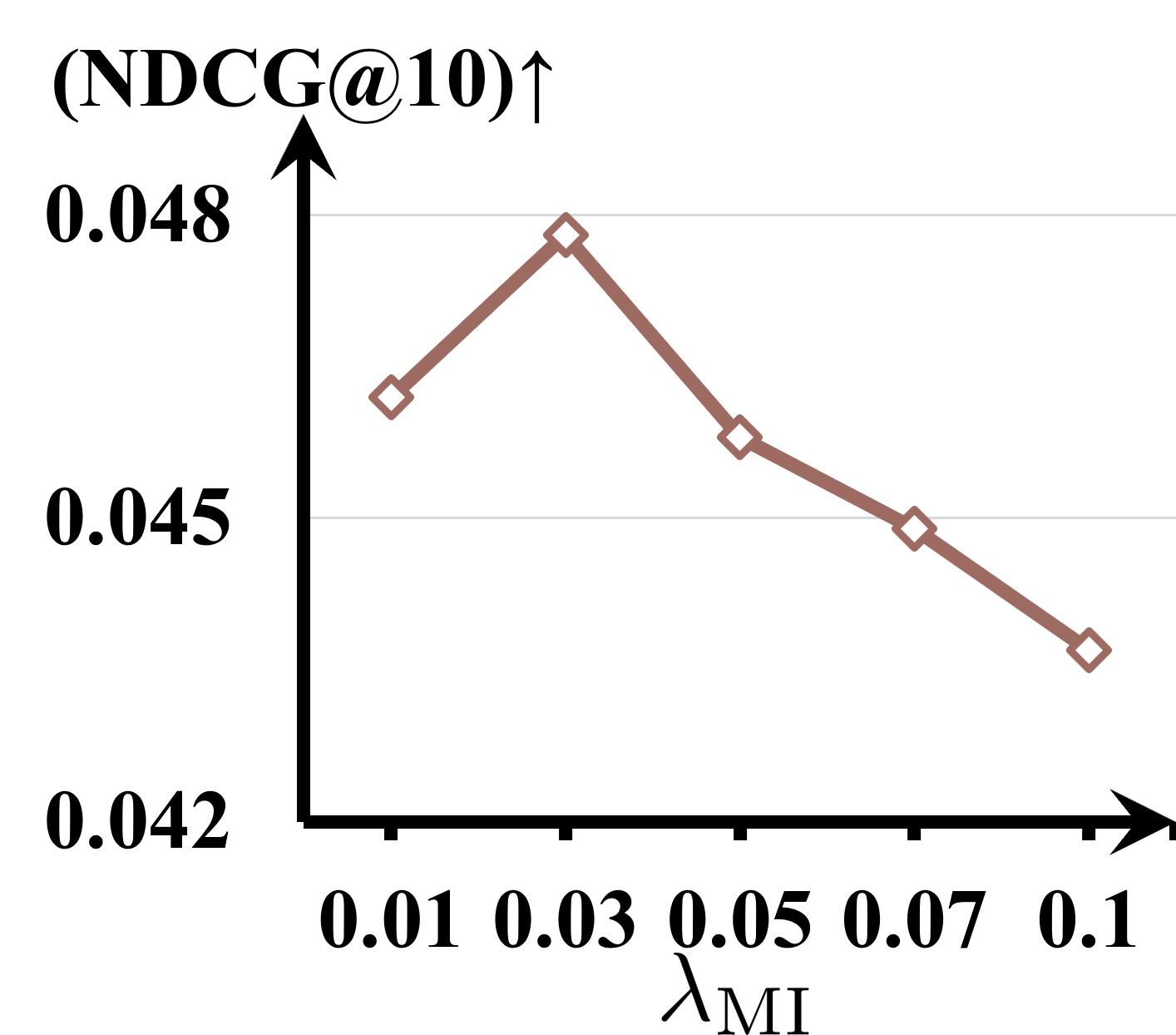} 
        \captionsetup{font={small,stretch=0.5}, skip=5pt,textfont=normalfont}
        \caption{}
        \label{fig:MI_beauty}
    \end{subfigure}
    \caption{Sensitivity analysis on Beauty.}
    \label{fig:sen_beauty}
\end{figure}

\begin{figure}[h]
    \centering
    \begin{subfigure}[b]{0.22\linewidth} 
        \centering
        \includegraphics[width=\linewidth]{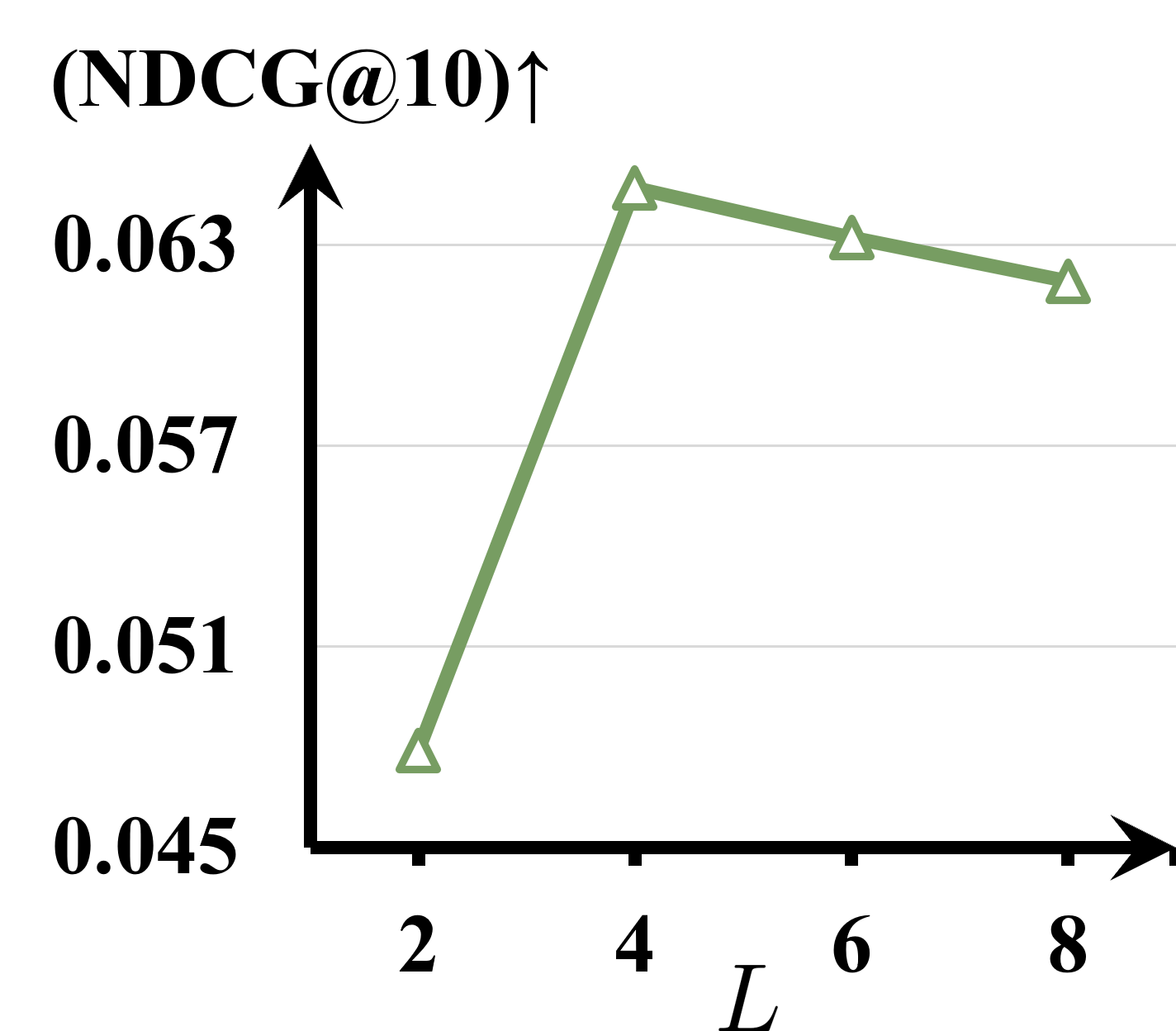} 
        \captionsetup{font={small,stretch=0.5}, skip=5pt,textfont=normalfont}
        \caption{}
        \label{fig:L_cellphones}
    \end{subfigure}
     \begin{subfigure}[b]{0.22\linewidth} 
        \centering
        \includegraphics[width=\linewidth]{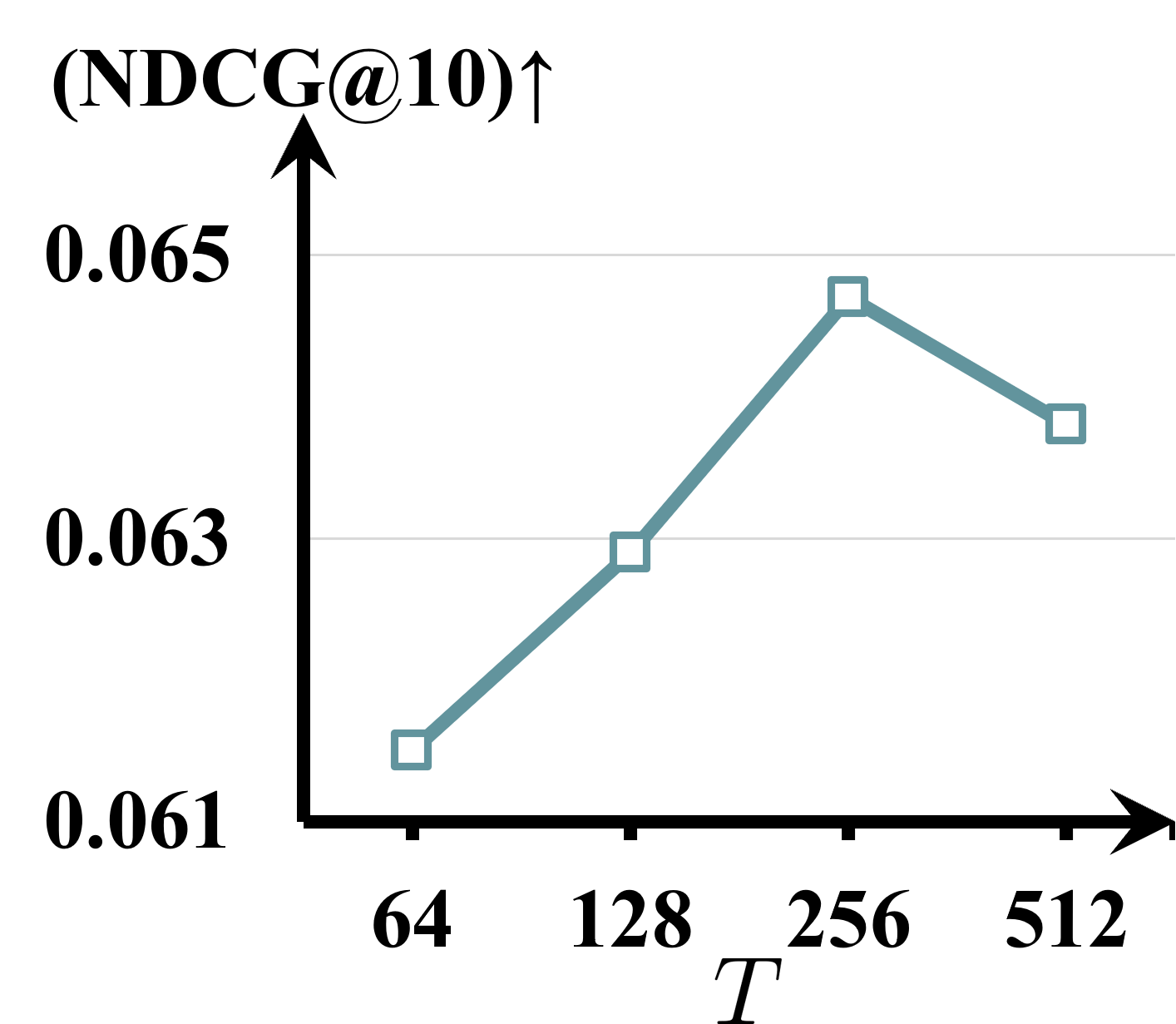} 
        \captionsetup{font={small,stretch=0.5}, skip=5pt,textfont=normalfont}
        \caption{}
        \label{fig:T_cellphones}
    \end{subfigure}
    \begin{subfigure}[b]{0.22\linewidth} 
        \centering
        \includegraphics[width=\linewidth]{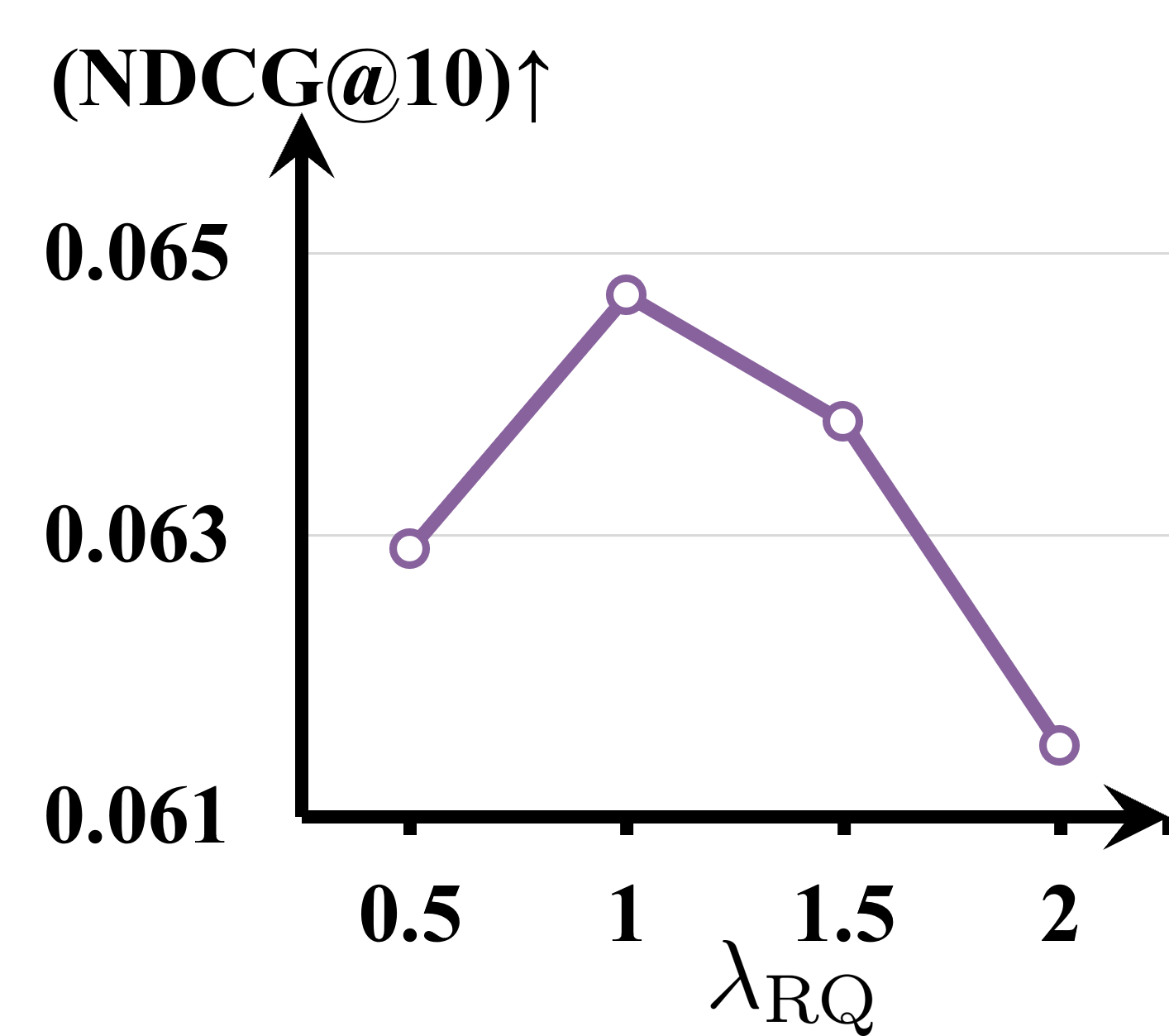} 
        \captionsetup{font={small,stretch=0.5}, skip=5pt,textfont=normalfont}
        \caption{}
        \label{fig:RQ_cellphones}
    \end{subfigure}
    \begin{subfigure}[b]{0.22\linewidth} 
        \centering
        \includegraphics[width=\linewidth]{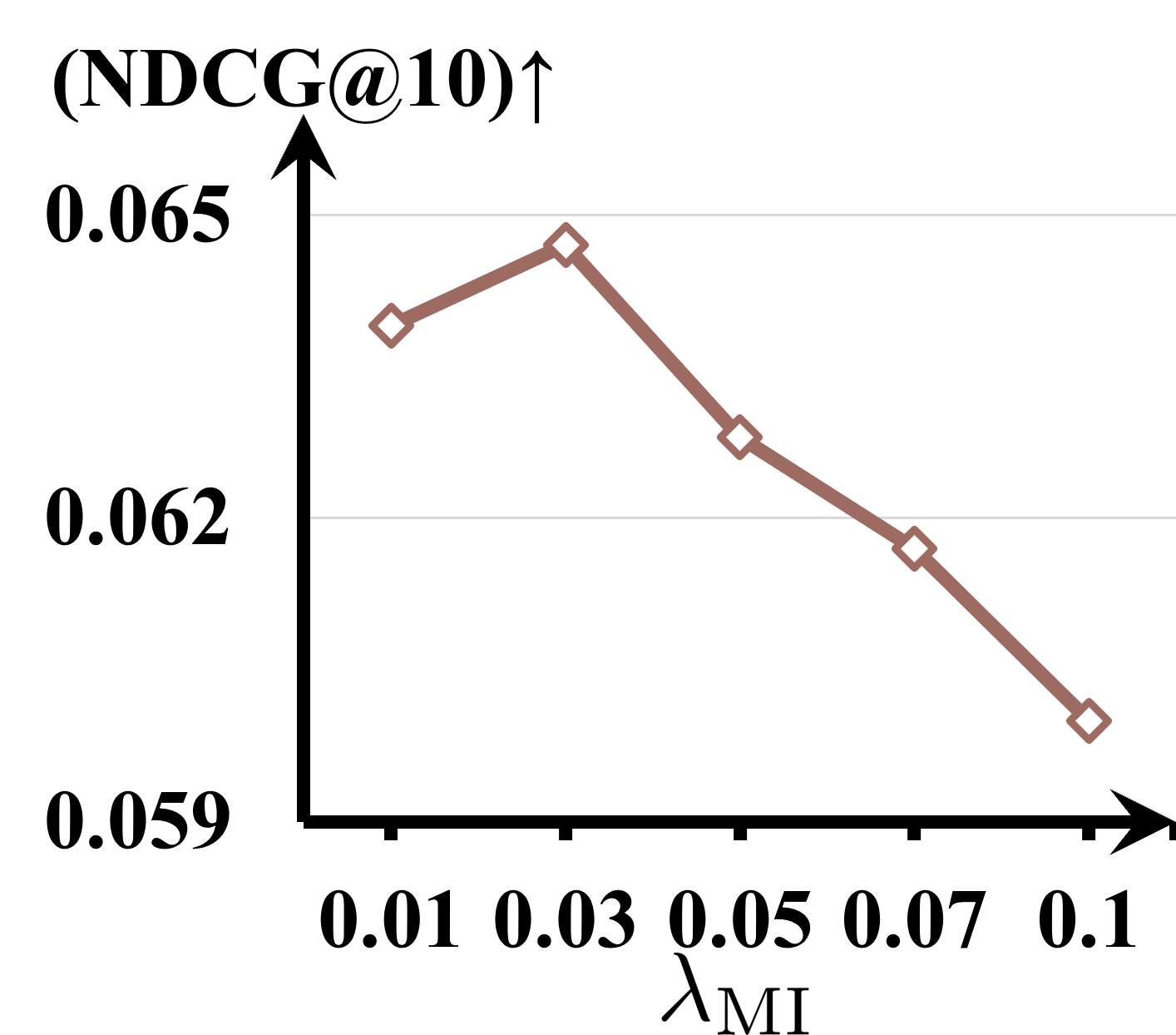} 
        \captionsetup{font={small,stretch=0.5}, skip=5pt,textfont=normalfont}
        \caption{}
        \label{fig:MI_cellphones}
    \end{subfigure}
    \caption{Sensitivity analysis on Cellphones.}
    \label{fig:sen_cellphones}
\end{figure}

\begin{figure}[th]
    \centering
    \begin{subfigure}[b]{0.22\linewidth} 
        \centering
        \includegraphics[width=\linewidth]{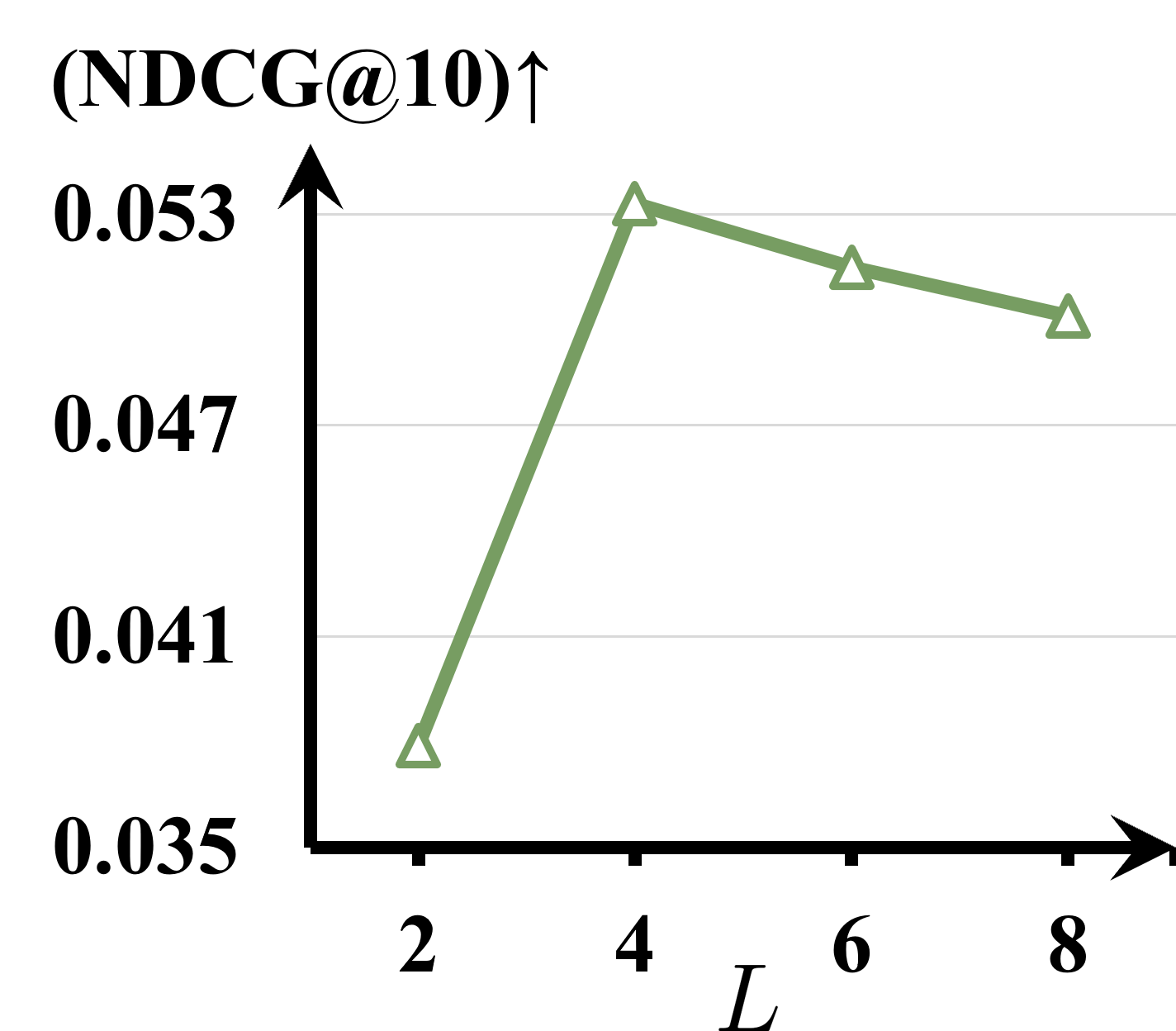} 
        \captionsetup{font={small,stretch=0.5}, skip=5pt,textfont=normalfont}
        \caption{}
        \label{fig:L_grocery}
    \end{subfigure}
     \begin{subfigure}[b]{0.22\linewidth} 
        \centering
        \includegraphics[width=\linewidth]{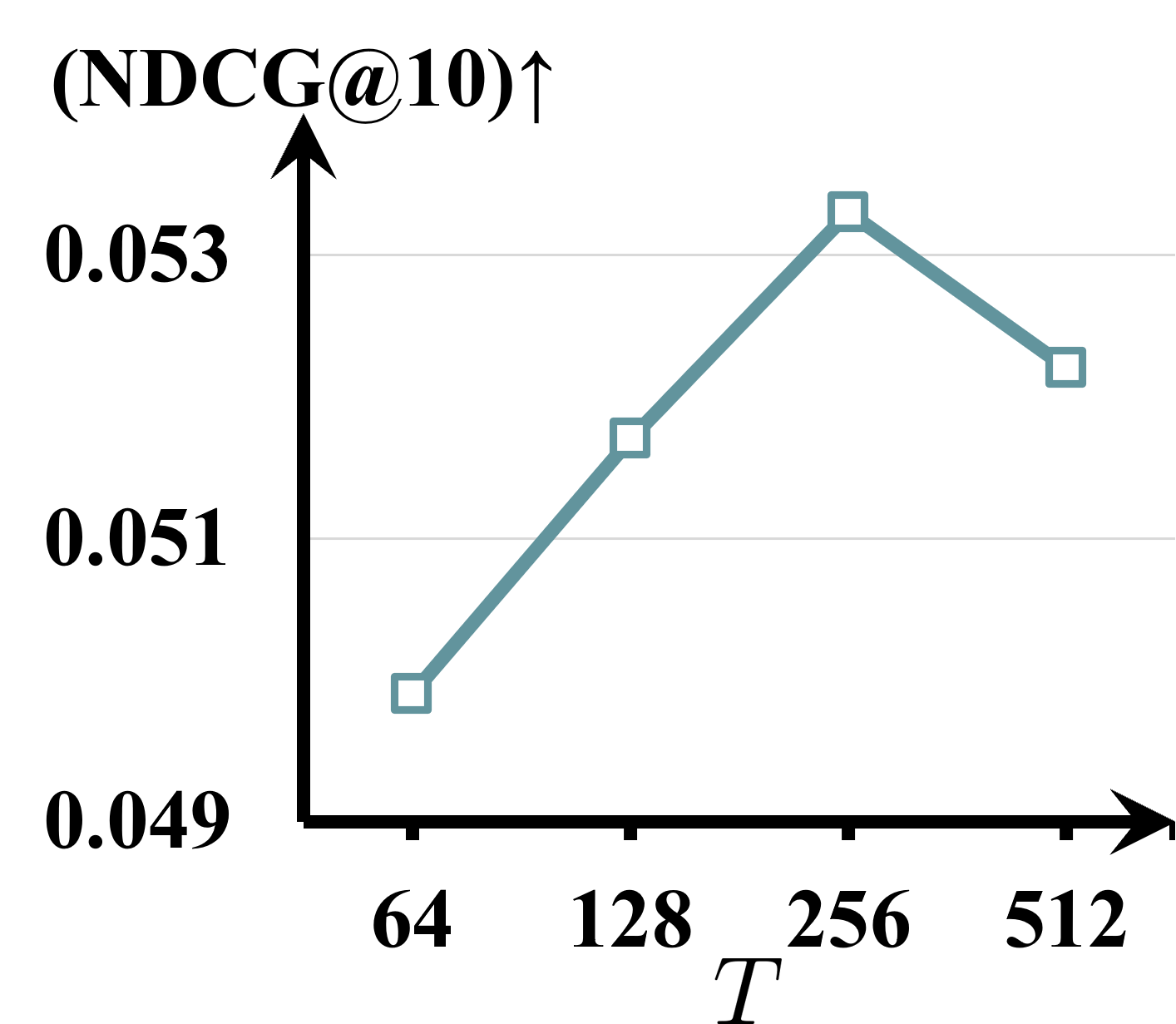} 
        \captionsetup{font={small,stretch=0.5}, skip=5pt,textfont=normalfont}
        \caption{}
        \label{fig:T_grocery}
    \end{subfigure}
    \begin{subfigure}[b]{0.22\linewidth} 
        \centering
        \includegraphics[width=\linewidth]{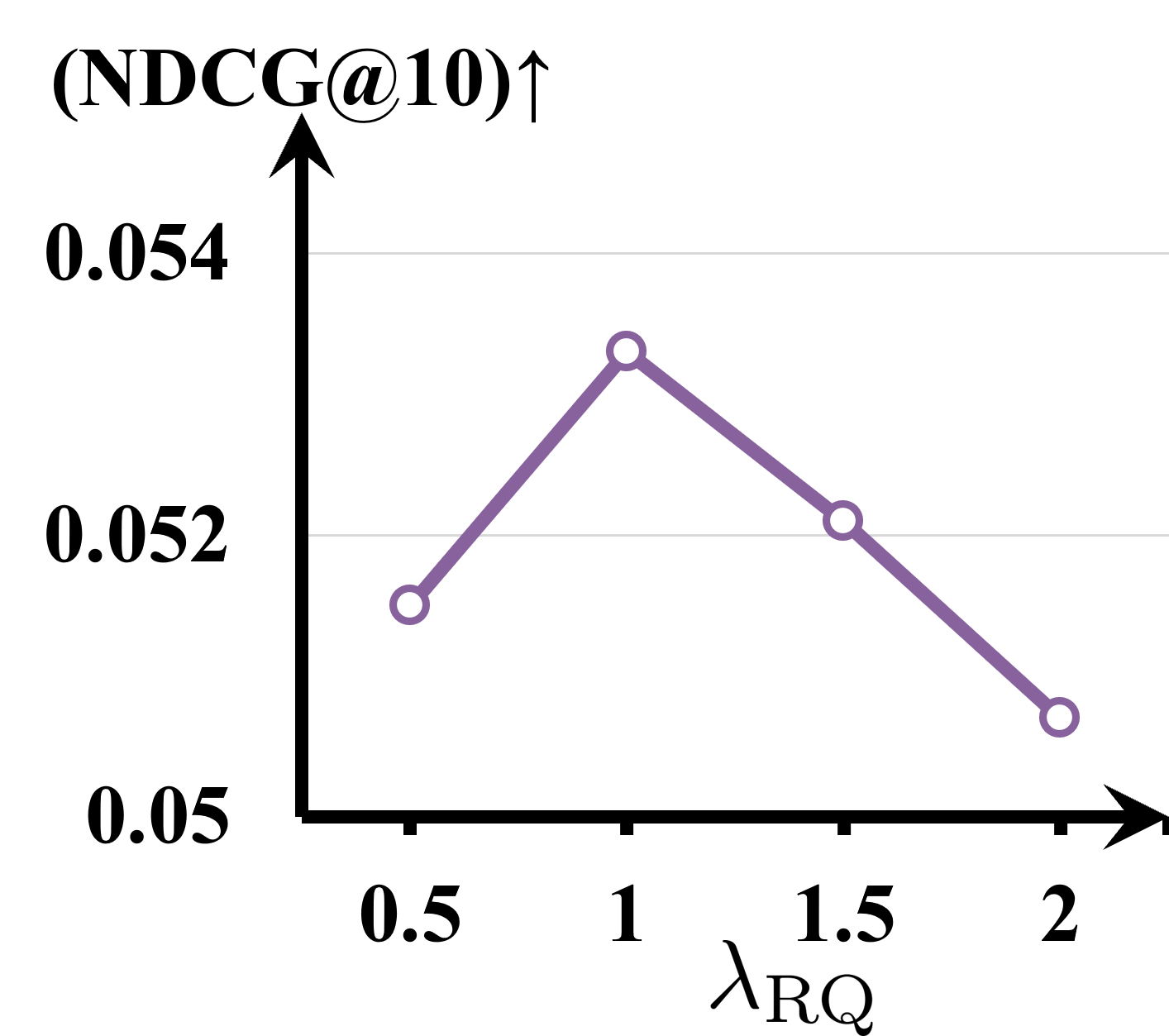} 
        \captionsetup{font={small,stretch=0.5}, skip=5pt,textfont=normalfont}
        \caption{}
        \label{fig:RQ_grocery}
    \end{subfigure}
    \begin{subfigure}[b]{0.22\linewidth} 
        \centering
        \includegraphics[width=\linewidth]{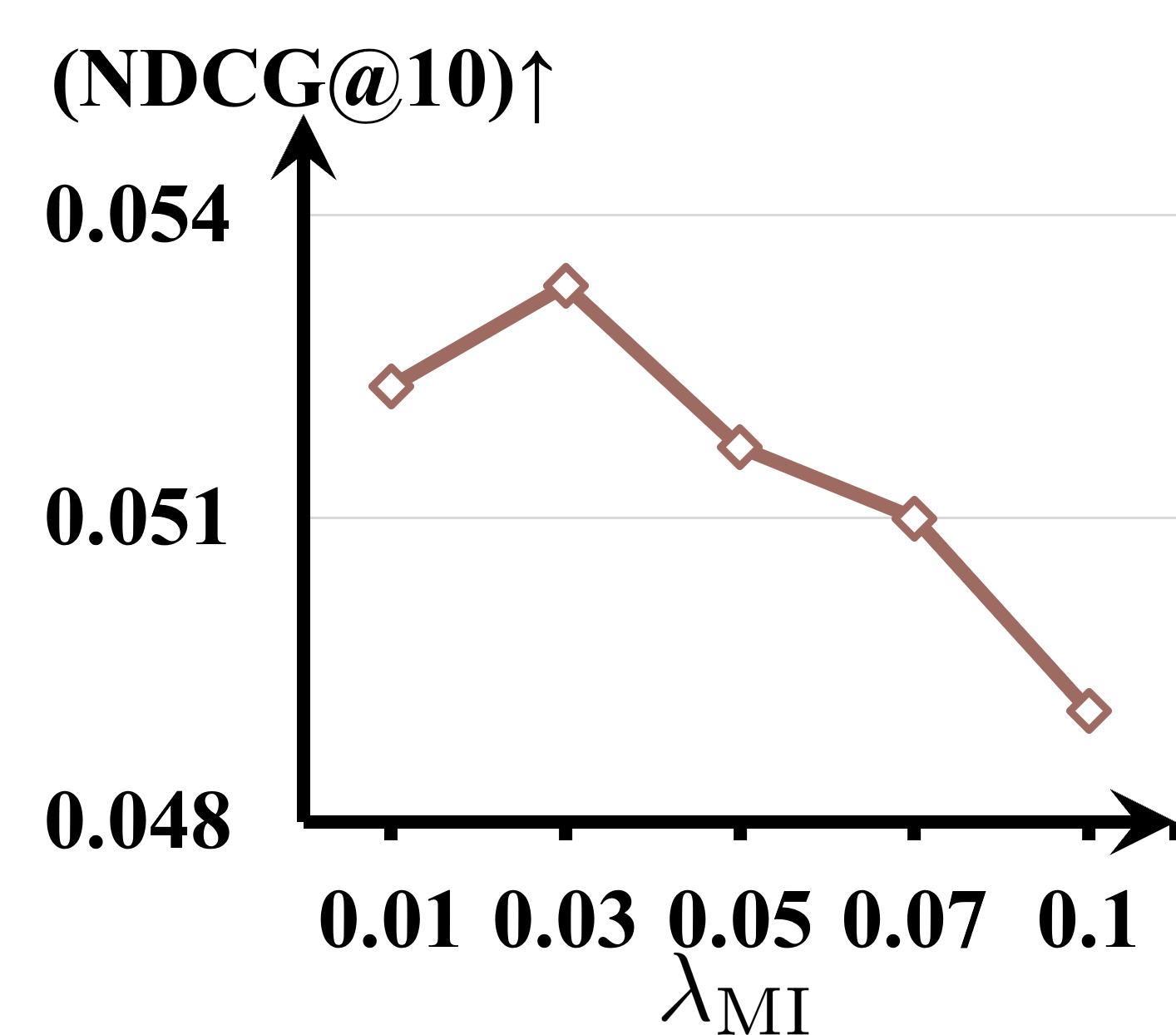} 
        \captionsetup{font={small,stretch=0.5}, skip=5pt,textfont=normalfont}
        \caption{}
        \label{fig:MI_grocery}
    \end{subfigure}
    \caption{Sensitivity analysis on Grocery.}
    \label{fig:sen_grocery}
\end{figure}

We analyze the sensitivity of \textsf{UniTok} to key hyperparameters, including the number of quantization levels $L$, codebook size $T$, and the loss weights $\lambda_\text{RQ}$ and $\lambda_\text{MI}$ in Eq. (11)  of the main manuscript. The results for the Beauty, Cellphones, and Grocery datasets are shown in Figures~\ref{fig:sen_beauty}, \ref{fig:sen_cellphones}, and \ref{fig:sen_grocery}, respectively. For clarity and focus, we present results on three representative domains that reflect trends consistent with those observed across the full range of evaluated datasets.

We first vary the number of quantization levels $L$ from 2 to 8 and report NDCG@10 in Figures~\ref{fig:L_beauty}, \ref{fig:L_cellphones}, and \ref{fig:L_grocery}. Performance improves as $L$ increases from 2 to 4, likely because longer sequences provide better capacity to capture fine-grained semantic information. However, further increasing $L$ beyond 4 leads to performance degradation, which can be attributed to error accumulation in longer autoregressive sequences during recommendations.

Next, we evaluate codebook sizes $T \in \left\{64,128,256,512\right\}$, with the results shown in Figures~\ref{fig:T_beauty}, \ref{fig:T_beauty}, \ref{fig:T_grocery}. Performance generally improves with larger codebooks, as they provide greater flexibility and token diversity distinguishing items. However, excessively enlarging the codebooks may lead to performance degradation. This may be due to the increased sensitivity to noise in item semantics, which can cause the model to overfit to spurious or less meaningful patterns.

We then examine the impact of the hyperparameter $\lambda_\text{RQ}$ in Eq.(11) of the main manuscript in terms of NDCG@10. As shown in Figure~\ref{fig:RQ_beauty}, \ref{fig:RQ_beauty}, \ref{fig:RQ_grocery}, performance peaks at $\lambda_\text{RQ}=1$ across datasets. The results suggest that setting $\lambda_\text{RQ}$ too high degrades performance, as it overemphasizes the residual quantizations that may be less relevant. Conversely, overly small $\lambda_\text{RQ}$ may underutilize codebook supervision. Notably, these findings highlight the importance of carefully tuning $\lambda_\text{RQ}$ to balance codebook-based identifiers and optimize performance.

Finally, we analyze the effect of the hyperparameter $\lambda_\text{MI}$ in Eq.(11) of the main manuscript in terms of NDCG@10. As shown in Figures~\ref{fig:MI_beauty}, \ref{fig:MI_cellphones}, and \ref{fig:MI_grocery}, the highest NDCG@10 is achieved at $\lambda_\text{MI}=0.03$. The results indicate that setting $\lambda_\text{MI}$ too high may degrade performance by overemphasizing mutual information calibration, potentially amplifying irrelevant variations. On the other hand, setting $\lambda_\text{MI}$ too low weakens the influence of mutual information preservation, limiting the semantic alignment of token representations. These findings highlight the importance of properly tuning $\lambda_\text{MI}$ to balance semantic preservation and generalization for optimal performance.





\subsection{8. Empirical Validation of Theoretical Claims}

We empirically validate the technical correctness and practical relevance of \textbf{Theorems 1--3}. Entropy analysis (Theorem 1) shows improved token entropy; quantization error comparison (Theorem 2) confirms a lower reconstruction error; and MI variance analysis (Theorem 3) demonstrates enhanced performance stability across domains.

\noindent
\textbf{Entropy analysis of token space (Supporting Theorem 1).} We compare the entropy of the token distributions produced by \textsf{UniTok} and codebook-based methods. Specifically, we analyze how much entropy gain is contributed by the router module in \textsf{UniTok} based on Eq. (12) of the main manuscript. The entropy is calculated based on the frequency of full token combinations across all items.

\begin{table}[h]
    \centering
    \begin{tabular}{lcc}
        \toprule
        \textbf{Method} & \textbf{Token Space Entropy} \\
        \midrule
        Codebook-based methods & 9.63 \\
        \textsf{UniTok} without the router & 9.63 \\
        \textsf{UniTok} (full) & \textbf{10.42} \\
        \bottomrule
    \end{tabular}
    \caption{Token space entropy comparison.}
    \label{tab:router_entropy}
\end{table}

\begin{figure}[h]
    \centering
    \includegraphics[width=0.4\columnwidth]{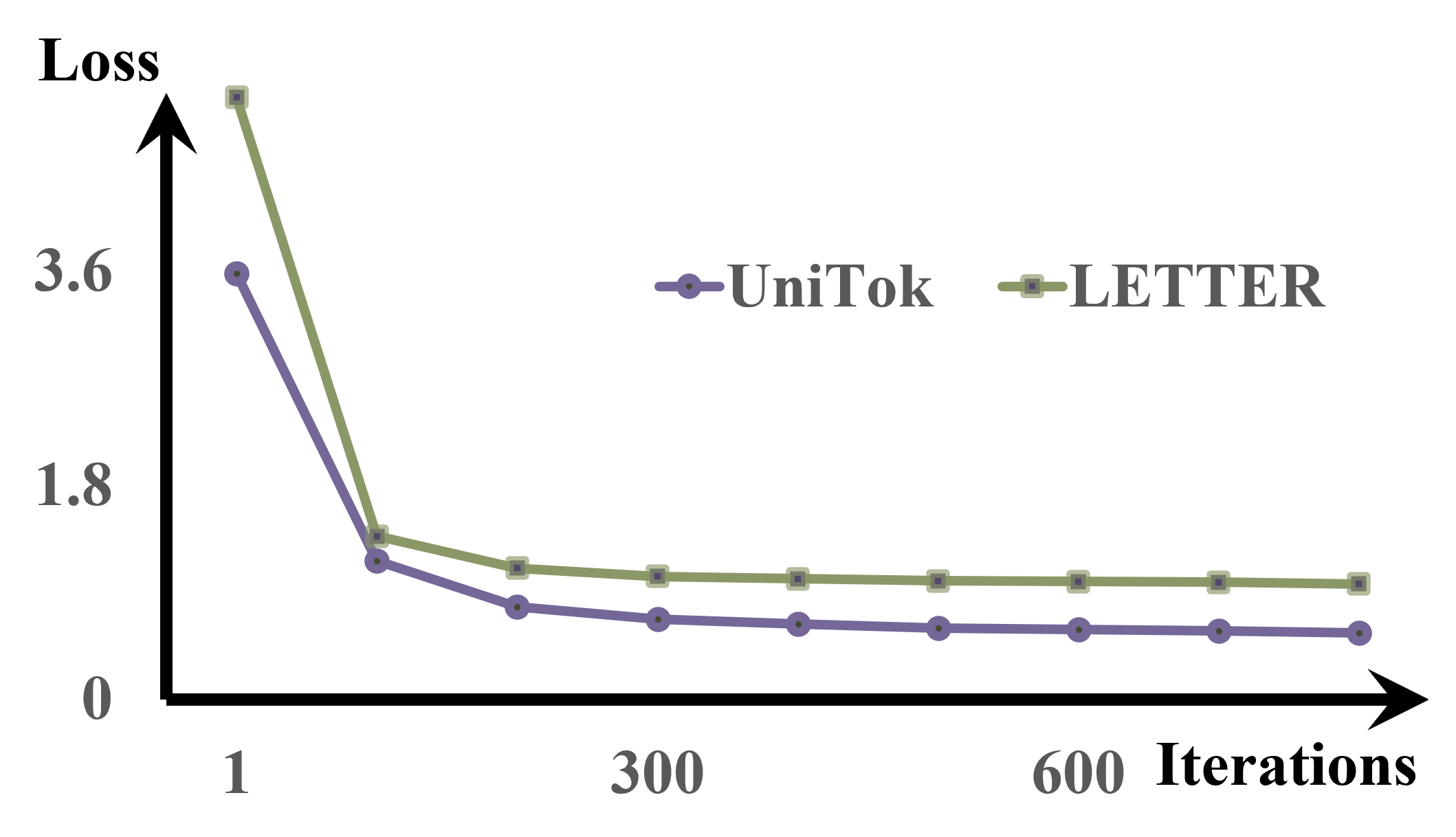}
    \caption{Comparison of residual quantization loss over training epochs between \textsf{UniTok} and LETTER.}
    \label{fig:theorem2}
\end{figure}

\begin{figure}[t]
    \centering
    \includegraphics[width=0.4\columnwidth]{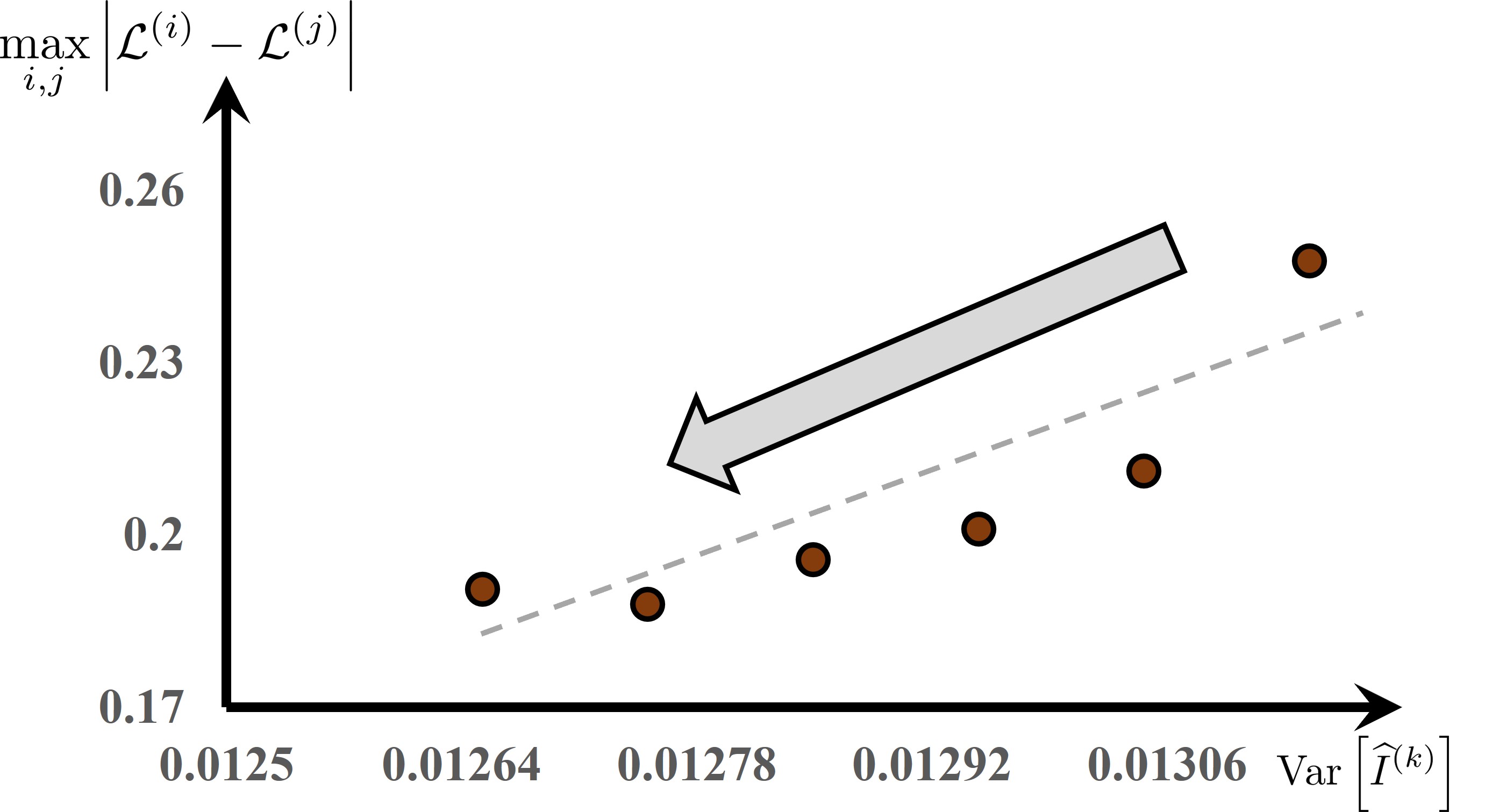}
    \caption{Relationship between MI variance and performance gap.}
    \label{fig:theorem3}
\end{figure}

As shown in Table~\ref{tab:router_entropy}, we observe that the router alone contributes an additional 0.79 Hartleys of entropy (measured using base-10 logarithm), expanding the capacity of the token space compared to standard codebook-based methods.

\noindent
\textbf{Quantization error comparison (Supporting Theorem 2).} To empirically support Theorem 2, which states that \textsf{UniTok} yields a lower expected quantization error than that of standard codebook-based methods, we compare their quantization losses on multi-domain setting. As shown in Figure~\ref{fig:theorem2}, \textsf{UniTok} consistently achieves a lower quantization error than the baseline case with a single set of codebooks. This confirms the theoretical insight that the TokenMoE architecture—by leveraging expert specialization—reduces a representation error and provides more precise item encoding. Notably, the expert-wise partitioning allows \textsf{UniTok} to partially compensate for domain-specific tokenization inaccuracies, leading to improved modeling fidelity in heterogeneous environments.

\noindent
\textbf{MI variance and performance stability (Supporting Theorem 3).} To validate Theorem 3, which posits that the variability in downstream performance across domains is upper-bounded by the variance of MI, we empirically examine the relationship between the variance of MI and the maximum difference in task loss across domains. Specifically, we compute the variance of MI estimates $\widehat{I}^{(k)}$ across domains, denoted as $\text{Var}\left[ \widehat{I}^{(k)} \right]$, and compare it against the observed performance variability $\max _{i,j} \left| {\mathcal{L}^{\left( i \right)}  - \mathcal{L}^{\left( j \right)} } \right| $ (Refer to Eq. (14) of the main manuscript). As shown in Figure~\ref{fig:theorem3}, we observe a strong positive correlation between $\sqrt{\text{Var}\left[ \widehat{I}^{(k)} \right]}$ and performance variability, consistent with the Lipschitz continuity assumption. This trend indicates that lower MI variance leads to more stable performance across domains, supporting our theoretical claim. These results suggest that MI serves as a reliable indicator of multi-domain representation consistency and can guide the design of more robust multi-domain recommendation systems.
\putbib[appendix]
\end{bibunit}

\end{document}